\documentclass{aa}  

\usepackage{graphicx}
\usepackage{appendix}
\usepackage{txfonts}
\usepackage{xcolor}
\usepackage{soul}
\usepackage{amsmath}
\usepackage{caption}
\usepackage{subcaption}
\usepackage{booktabs}
\usepackage{enumitem}
\usepackage{natbib}
\usepackage{csquotes}
\usepackage{float}
\usepackage{makecell}
\usepackage{booktabs}
\usepackage{multirow}
\usepackage{xspace}
\usepackage[colorlinks=true, citecolor=violet, linkcolor=red, urlcolor=blue]{hyperref}
\begin{document} 

\newcommand{\bpop}{\textsc{B-pop~}}
\newcommand{\dragon}{\textsc{DRAGON-II}}
\newcommand{\globular}{GCs\xspace}
\newcommand{\nuclear}{NSCs\xspace}
\newcommand{\young}{YCs\xspace}
\newcommand{\msun}{{\rm M}_\odot}

\newcommand{\red}[1]{\textcolor{red}{#1}} 
\newcommand{\blue}[1]{\textcolor{blue}{#1}} 
\newcommand{\green}[1]{\textcolor{Green}{#1}} 
\newcommand{\poptxt}[1]{\textbf{\textcolor{brown}{#1}}} 
\newcommand{\lav}[1]{\textcolor{blue}{\sf{[LP: #1]}}}
\newcommand{\CU}[1]{\textcolor{orange}{\sf{[CU: #1]}}}

\newcommand{\orcidicon}[1]{\href{https://orcid.org/#1}{\includegraphics[width=11pt]{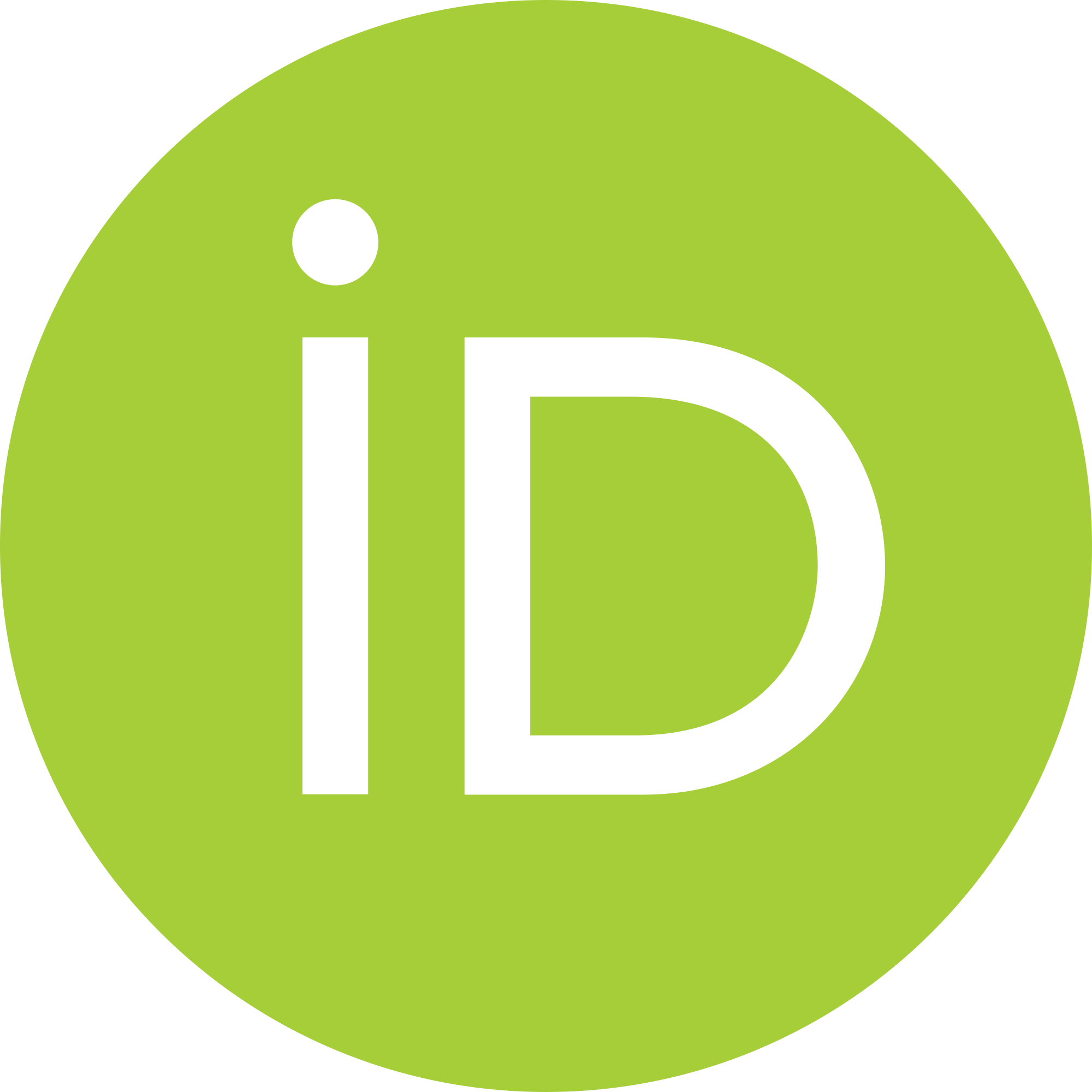}}}
\newcommand{\orcid}[1]{\href{https://orcid.org/#1}{\protect\orcidicon{#1}}}


   \title{\textit{Seeds to success:    
   }\\ growing heavy black holes in dense star clusters}


   \author{Lavinia Paiella
          \inst{1,2}   \orcid{0009-0001-7605-991X}
          \and
          Manuel Arca Sedda\inst{1,2}  \orcid{0000-0002-3987-0519} 
          \and
          Benedetta Mestichelli\inst{1,2,3}
          \orcid{0009-0002-1705-4729} 
          \and
          Cristiano Ugolini \inst{1, 2}
          \orcid{0009-0005-9890-4722}
          }
   \authorrunning{L. Paiella et al.}
   \institute{$^1$Gran Sasso Science Institute (GSSI), L'Aquila (Italy), Viale Francesco Crispi 7\\
            $^2$ INFN, Laboratori Nazionali del Gran Sasso, 67100 Assergi, Italy\\
            $^3$ Universit\"at Heidelberg, Zentrum f\"ur Astronomie (ZAH), Institut f\"ur Theoretische Astrophysik, Albert Ueberle Str. 2, 69120, Heidelberg, Germany
             \\\\
            \email{lavinia.paiella@gssi.it}\\
              \email{manuel.arcasedda@gssi.it}
             }


 
  \abstract
   {
   The observational dearth of black holes (BHs) with masses between $\sim$100 and 100,000 $\msun$ raises questions about the nature of intermediate-mass black holes (IMBHs). 
   Proposed formation channels for IMBHs include runaway stellar collisions and repeated binary BH (BBH) mergers driven by dynamical interactions in stellar clusters, but the formation efficiency of these processes and the associated IMBH occupation fraction are largely unconstrained.}
   {In this work, we study IMBH formation via both mechanisms in young, globular, and nuclear star clusters. We carry out a comprehensive investigation of IMBH formation efficiency by exploring the impact of different seeding models and star cluster formation histories.}
   {We employ a new version of the \bpop population synthesis code, able to model several seeding mechanisms as well as hierarchical BBH mergers. }
   {We quantify the efficiency of IMBH production across different cluster families, and estimate the fraction of BBH mergers involving an IMBH primary. Comparison with low-redshift IMBH candidates suggests that, depending on the seeding mechanism, stellar collisions can play a pivotal role in explaining potential IMBHs in local globular clusters.}
   {Our simulations highlight stellar collisions as the primary IMBH formation channel across a wide range of cluster types. They further suggest that wandering IMBHs may populate Milky Way-like galaxies and that correlations between cluster and IMBH masses can help distinguish the origins of Galactic globular clusters.}


   \maketitle
%




\section*{Introduction}
Black holes (BHs) can be broadly classified into two main categories based on their mass.
Stellar-mass BHs lie at the low end of the mass scale, with masses ranging from about $3-5 \, \msun$ up to roughly 100 $\msun$. These BHs form from the collapse of massive stars with initial masses greater than approximately $20\, \msun$ \citep[see e.g.][]{vink_2001, woosley_2002, fryer_2012,spera_2015, vink_2021, ugolini_2025}.
At the opposite extreme are supermassive BHs (SMBHs), which have typical masses exceeding $\sim 10^6\,\msun$. SMBHs are commonly found at the centers of galaxies and are believed to grow through complex evolutionary processes, starting from lighter BH "seeds" that increase in mass via various mechanisms \citep[see e.g.][]{volonteri_2021, lupi_MBHs_review_2023}.
These seeds, which often represents intermediate states between the two BH mass categories, fall in the mass range between $\sim 10^2$ and  $10^5\,\msun$ and belong to the elusive family of intermediate mass black holes (IMBHs).
IMBHs 
have been extensively studied through a variety of observational channels, including searches for potential radiative accretion signatures \citep{maccarone_radio_2008,tremou_2018,tde_lin_2018,tiengo_2022}, integral field spectroscopy and astrometric measurements of stellar motion \citep{gebhardt_g1_2005, noyola_omegacen_2010,vandermarel_omegacen_2010,lutzgendorf_2013,lanzoni_2013,kamann_2014,kamann_2016,pechetti_2022,haberle_omegacen_2024}, millisecond pulsars timing \citep{damico_2002,colpi_2003,ferraro_2003,kiziltan_2017,perera_2017,gieles_2018,banares_2025}, and gravitational-wave (GW) detections \citep{gw190521_letter_2020, gw231123_2025}. 
Despite the increasing observational evidence, their existence is often difficult to confirm due to the absence of a definitive observational signature and the possibility of their effects to be mimicked by other astrophysical systems or processes. Investigating IMBH formation in different stellar environments, and its potential smoking guns, is, therefore, fundamental to understand the nature of these objects and their possible links to stellar and SMBHs.

One promising route for IMBH formation is through dynamical assembly in a dense star cluster. Within this formation scenario, several mechanisms can play a role in the production and growth of the IMBH. 
Among them are the formation of a very massive star (VMS) from repeated stellar collisions that either collapses into an IMBH or accretes most of its mass on a stellar BH, repeated mergers among BHs and/or stars, or, finally, multiple tidal disruption events on stellar BHs \citep{millerhamilton2002, portegies_zwart_runaway_2002,  giersz_2015, dragonII_II_2023, rizzuto_2023_tdes}. 
Repeated stellar collisions may occur in light ($\lesssim 10^6\,\msun $) and compact ($\lesssim 0.1-1\,\rm pc$) star clusters, where core collapse sets in before massive stars evolve in compact objects. In these clusters, the density peak reached at collapse can trigger the onset of collisions among massive stars which can proceed in a runaway fashion
, hence the term runaway stellar collisions \citep{portegies_zwart_runaway_2002}. The resulting VMS can reach a mass of $\sim 0.1-1 \%$ of the cluster’s initial mass, eventually collapsing into an IMBH. 
Conversely, hierarchical binary BH (BBH) mergers \citep{millerhamilton2002} can trigger the build-up of an IMBH starting from a stellar BH progenitor, or yield the further growth of a seed formed from stellar collisions.
This process is expected to take place preferentially 
in massive, compact clusters with large escape velocities ($\gtrsim$ a few hundreds km/s).

Historically, these two fundamental formation channels have been extensively studied through direct $N$-body and Monte Carlo simulations 
(see e.g. \citealp{giersz_2015, mapelli_2016, kremer_2020,dicarlo_2021, rizzuto_ycs_2021, dragonII_II_2023, vergara_2024, barber_antonini_2024, gonzalez_2024, frost_rantala_2024, vergara_2025}). However, the computational costs of both techniques restrain the 
cluster parameter space that can be investigated (e.g. cluster structure, metallicity, formation history).
An appealing alternative that became quite popular in the last few years is represented by semi-analytic codes, which encode simplified recipes to model BH dynamics and significantly reduce the required simulation time 
\citep{antoninirasio2016,arcasedda_benacquista_2019,antonini_gieles_2020,bpop_2020,fast_cluster_2021,arcasedda_isolated_2023, rapster_2024, torniamenti_2024}.

In this work we employ the latest version of 
the semi-analytic population synthesis code \bpop \citep{arcasedda_benacquista_2019, bpop_2020,arcasedda_isolated_2023} to explore possible relations among different IMBH formation scenarios and the properties of their nursing environments. 
In Section \ref{sec::methods} we present the main features of \bpop and the different models we investigate for the seeding and the clusters formation history. In Section \ref{sec:results} we briefly overview the efficiency of different IMBH formation and growth mechanisms based on the initial properties of the host clusters. Moreover, we present the results of our simulations, focusing on the emergence of IMBHs at low redshifts and their similarities to potential observed candidates. In Section \ref{sec::discussion} we discuss some interesting implications of our findings for different families of clusters. Finally, in Section \ref{sec:conclusions} we summarize our main conclusions.
\section{Methods}
\label{sec::methods}
We study the dynamical formation of IMBHs in star clusters under different seeding prescriptions and cluster formation histories using the semi-analytic population synthesis code \bpop \citep{arcasedda_benacquista_2019, bpop_2020, arcasedda_isolated_2023}. The code constructs a synthetic Universe filled with BBHs and characterized by different cosmic star formation rates for each cluster type, a redshift-dependent metallicity distribution, a reliable time evolution for the structural properties of star clusters, and a flexible environment where to test different assumptions for the BHs initial properties. A detailed description of the code can be found in the companion work (Arca Sedda et al., in prep.), our previous works \citep{arcasedda_isolated_2023} and in Appendix \ref{app::theory}. 
\subsection{The \bpop code}
\label{sec::bpop}
We use \bpop to simulate $10^8$ BBHs evenly split among young (\young), globular (\globular) and nuclear star clusters (\nuclear). In Table \ref{tab:simuparams} we summarize the main parameters of our simulations which we describe in more details in the following along with the code's main features.
\subsubsection{Clusters}
The initial mass and half-mass radius distributions of each cluster family are tuned on observations at low redshifts. \young are modeled to match observations in the Milky Way and the Local Volume \citep{pz_ecology_2006, gatto_ripepi_2021}. While the half-mass radius of these clusters is well constrained, dynamical mass estimates are available for only a few systems. Therefore, \bpop samples the half-mass radii of \young directly from the observed distribution, while their masses are drawn from the \globular mass distribution, reduced by 2.5 dex, in line with observations of \young in the Milky Way and its satellites \citep{ycs_review_PZ_2010}. The masses and radii of \globular are extracted from the observed distributions of galactic \globular \citep{harris_1996, baumgardt_2018}, while the properties of \nuclear are tailored to match observations of these clusters in local galaxies \citep{georgiev_2016_nuclear}.

\begin{figure}[h]
    \centering
    \includegraphics[width=
0.9\columnwidth]{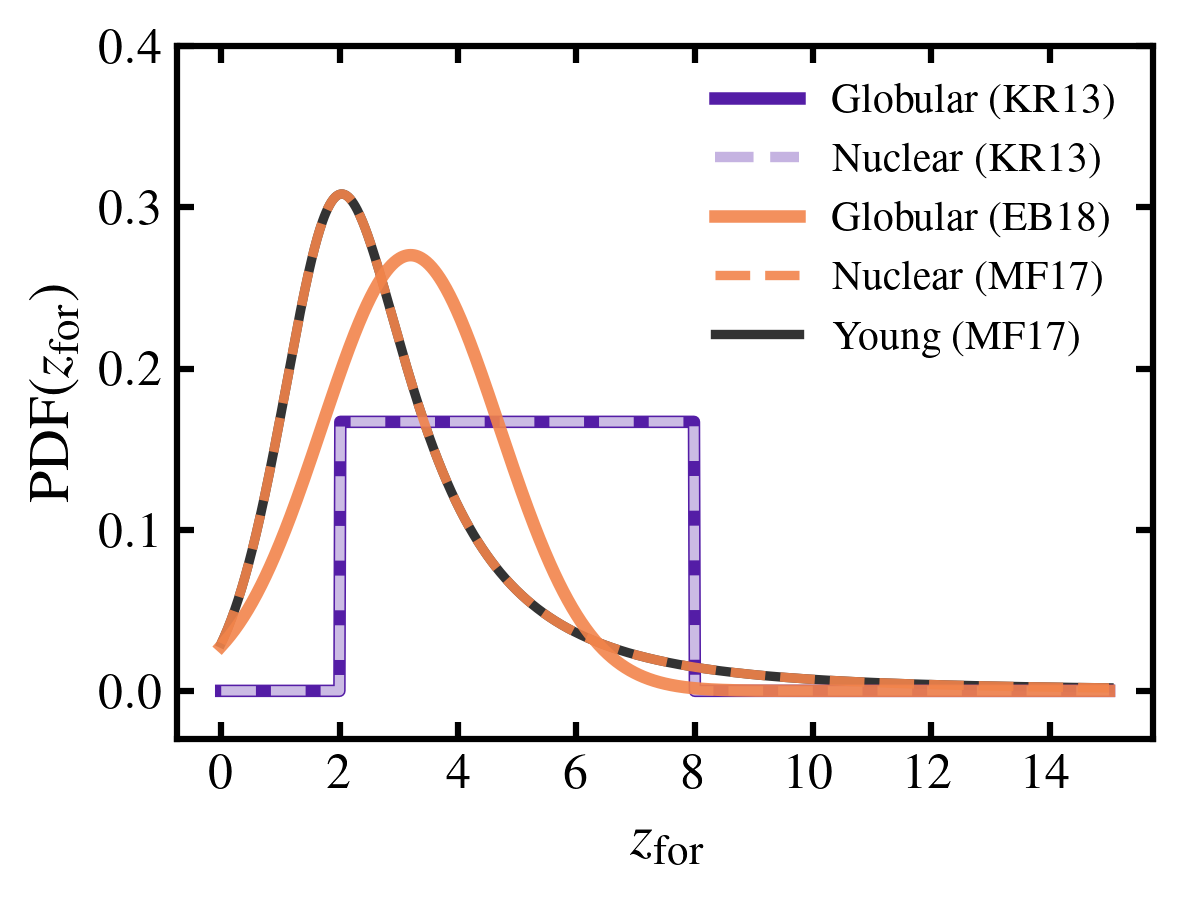}
    \caption{\small Initial distribution of clusters' formation redshift. Different colors refer to different assumptions on the initial formation histories of \globular (solid lines) and \nuclear (dashed lines). The initial redshift distribution of \young is fixed in all models (grey solid line). Distributions are conveniently scaled such that the area subtended by the curve is 1.}
    \label{fig::sfr}
\end{figure}

In Figure \ref{fig::sfr} we show the different cluster formation redshift distributions used in this work  \citep[see also Figure 4 in our previous paper, ][]{arcasedda_isolated_2023}.
In all our simulations, we assume that \young follow the star formation history of their host galaxies, sampling their formation redshifts from the star formation rate (SFR) of \citet[hereafter MF17]{madaufragos2017}. 
For \globular and \nuclear we consider two options. In the first, we adopt a flat redshift distribution in the range $z \in [2, 8]$, as suggested by observations of extragalactic \globular by \citet[hereafter KR13]{katz_two_2013}. In the second, we sample the \nuclear formation redshifts from the SFR of MF17 and the \globular formation redshifts from the distribution obtained in the cosmological simulations of \citet[hereafter EB18]{elbadry_2019}.

Regardless of the formation scenario, we assume 
that the mean metallicity of stellar progenitors evolves with redshift as $\log \langle Z / Z_\odot \rangle \simeq 0.153 - 0.074 \ z^{1.34}$ (MF17). For each value of $Z$, we assume the metallicity distribution is spread around the mean value following a log-normal distribution peaked at $\log \langle Z / Z_\odot \rangle $ and with dispersion $\sigma_{Z} = 0.2$ \citep{bavera_2020, santoliquido_2022, mapelli_2022}.

\subsubsection{Black holes}
In \bpop, the masses of BHs formed from single stars or in stellar binaries are drawn from catalogs generated using external stellar evolution tools. To ensure continuity with our previous works \citep{arcasedda_isolated_2023}, we adopt catalogs generated with \textsc{MOBSE} \citep{giacobbo_mapelli_2018, giacobbo_mapelli_2020} covering metallicities in the range $Z \in [0.0002, 0.02]$. 
In particular, BH masses can be extract from either (i) a simple single stellar BH mass (SSBM) spectrum, corresponding to BHs formed from single stars, or (ii) a mixed single BH mass (MSBM) spectrum, which represents the population of BHs formed from binary stellar systems \citep{arcasedda_isolated_2023}. 
In all our runs BHs have $50\%$ probability to have a mass sampled from the MSBM or SSBM spectrum. Note that this subtly implies the inclusion in our dynamical models of effects related to primordial binary evolution, e.g. star mergers or accretion episodes \citep{arcasedda_isolated_2023}, a feature that may have a significant impact on the emergence of single and double IMBHs with a mass $<300$ M$_\odot$ (see e.g. \cite{paiella_letter_2025}). In both mass spectra the initial mass of a stellar-mass BH never exceeds the minimum IMBH mass, i.e. $100 \,\msun$. Nonetheless, additional features implemented in \bpop allow the user to assign BH masses in and beyond the pair instability supernova (PISN) range \citep{woosley_2017}, e.g. assuming that their progenitors formed from one or multiple stellar collisions. We exploit this feature to explore the impact of different prescriptions for the early seeding of IMBHs, as detailed in Section \ref{subsubsec::runaway}, \ref{subsubsec::seedingindense} and \ref{subsec::models}.

In our runs a BH (IMBH) can grow dynamically via repeated BBH mergers. After each merger the retention of the remnant is ensured only if its relativistic kick is lower than the escape velocity of the environment (see more on this in Section \ref{subsubsec::hierarchical}). This kick, which arises from asymmetries in GW emission prior to the merger, is primarily influenced by the mass ratio ($q = m_1 / m_2 < 1$) of the BBH and the magnitude and relative alignment of the initial spins \citep{campanelli_2007, lousto_2008, lousto_2012}. 
In our simulations 
we always assume a BH (IMBH) to pair up with BHs which have not undergone any previous merger. 
While neglecting mergers with higher-generation BHs is a simplification, multiple IMBHs are unlikely to form and grow within the same environment \citep[see e.g.][]{gurkan_2006, giersz_2015, mapelli_2016, kovetz_2018, rodriguez_2019, anagnostou_2020, gonzalez_2021, dragonII_II_2023, rantala2025rapidchannelcollisionalformation}, except for clusters with either a large fraction of primordial binaries, extreme densities, or undergoing a hierarchical formation \citep[see e.g.][]{gurkan_2006,Maliszewski_2022, rantala2025rapidchannelcollisionalformation}.
A prescription for the pairing up of BHs from different generation is under construction for the future public release of the code (Ugolini et al., in prep).

Concerning the clusters' BHs initial spins
we always assume 
them to be distributed isotropically and with magnitudes following a Maxwellian distribution with dispersion $\sigma_\chi = 0.2$. This distribution is consistent with the inferred spin distribution of merging BBHs observed by the LVK collaboration up to the O3 run \citep{abbott_2023b}. 
\begin{table*}[h]
    \centering
    \small
    \renewcommand{\arraystretch}{1.5} 
    \setlength{\tabcolsep}{8pt} 
    \begin{tabular}{p{0.3\columnwidth}p{0.35\columnwidth}p{0.9\columnwidth}}
    \toprule
      Symbol &   Value &  Description \\
    \midrule
    \midrule
        $N_{\rm tot}$ & $ 10^8$ & Number of simulated BHs. \\
        $P_{\rm mix}$ & 0.5 & Probability of a BH to be extracted from a MSBH spectrum. \\
        PDF$(\chi)$   &  Maxwellian, $\sigma = 0.2$ &    Initials spin distribution. \\
        $f_{\rm \young},f_{\rm \globular}, f_{\rm \nuclear}$& 1/3, 1/3, 1/3 & Fraction of \young, \globular and \nuclear. \\
        PDF$(z_{\rm for})$ & MF17, KR13, KR13 \textit{or} MF17, EB18, MF17 & Formation redshifts distribution of \young, \globular and \nuclear. \\
    \bottomrule
    \end{tabular}
    \caption{\small Main \bpop input parameters in our simulations (Column 1) along with their assigned value (Column 2) and a short description (Column 3).  We list the total number of BHs simulated per model ($N_{\rm tot}$),
    the probability of a BH to be sampled from a MSBM spectrum ($P_{\rm mix}$), the initial spins distribution (PDF$(\chi)$), the fraction of \young, \globular and \nuclear ($f_{\rm \young}$, $f_{\rm \globular}, f_{\rm \nuclear}$) and their formation redshift distributions (PDF$(z_{\rm for})$). 
    }
    \label{tab:simuparams}
\end{table*}


\subsection{IMBH formation channels}
\label{subsec::theoframework}
In the following, we briefly describe the processes regulating IMBH seeding and growth considered in this work, namely the stellar collision scenario and the hierarchical BBH merger scenario. Note that the two processes are not mutually exclusive, e.g. an IMBH formed from the collision of a few massive stars can indeed further grow by merging with stellar BHs. For the sake of conciseness, we provide a more in-depth discussion about these mechanisms in Appendix \ref{app::theory}.
\subsection{Cluster relaxation and core-collapse}
Any gravitationally bound system undergoes dynamical relaxation driven by the continuous dynamical interactions among its constituents. In a star cluster, this process occurs over a timescale \citep{spitzer_1969, binney_tremaine_2008}:
\begin{equation}
    t_{\rm rel} = 4.2\,\mathrm{Gyr}\,\left( \frac{15}{\ln \Lambda_{\rm c}}\right)\,\left( \frac{R_{\rm h , cl}}{4 \rm\,pc}\right)\,\sqrt{\frac{M_{\rm cl}}{10^7 \, \msun}}\,,
    \label{eq::trel}
\end{equation}
called half-mass relaxation time, which depends 
on the cluster mass ($ M_{\rm cl}$) and the radius within which half of the cluster mass is contained, i.e. the half-mass radius ($R_{\rm h, cl}$). 
Relaxation causes light stars to diffuse outward, carrying kinetic energy away from the core. This energy loss makes the core contract, which in turn increases the diffusion, creating a loop that leads the core to collapse.
This phenomenon is commonly referred to as gravothermal instability \citep{spitzer_1969}.

The contraction is finally reversed by the formation of hard binaries, that transfer kinetic energy to other core members through dynamical interactions and leads the core to bounce back. 
The core collapse process is, therefore, fundamentally governed by dynamical interactions between light and massive stars. These interactions also promote the segregation of the most massive stars in the cluster center on a timescale \citep{spitzerhart_1971, pz_2004}:
\begin{equation}
    t_{\rm s}(m) = \frac{\bar m_*}{m} \, \frac{0.138\,N_*}{\log(\Lambda_{\rm c})} \, \left(\frac{R_{\rm h, cl}^3}{ G M_{\rm cl}}\right)^{1/2}\,, 
    \label{eq::tseg}
\end{equation}
where $m$ is the mass of the sinking star, $\bar m_*$ is the average mass of a star in the cluster and $N_* \sim M_{\rm cl}/\bar m_*$ is the average total number of stars assuming an initial cluster mass $M_{\rm cl}$. For our simulations we fix $\bar m_*= 0.7\,\msun$ and assume that the core collapse happens on a timescale equal to the segregation time of the most massive stars in the cluster population, i.e. for $t_{\rm cc} \sim t_{\rm s} (150 \, \msun)$. Moreover, we define a lower bound for the core collapse timescales of $t_{\rm cc} \sim 0.2\,t_{\rm rel}$ \citep[see also][]{portegies_zwart_runaway_2002}.

\subsubsection{Runaway stellar collisions and IMBH seeds}
\label{subsubsec::runaway}
If the core-collapse happens on timescales shorter than $\sim 3-5\,\rm Myr$, which is the typical lifetime of BH massive stellar progenitors, it can trigger a chain of runaway stellar collisions and the formation of a VMS \citep{portegies_zwart_runaway_2002} that may collapse into an IMBH. 
\\
The maximum mass of the VMS can be conveniently re-written as a function of the initial cluster mass:
\begin{equation}
     m_{\rm VMS} = m_{\rm seed} + 800\,\msun \,\left( \frac{f_{\rm c}}{0.2} \right)  \left( \frac{M_{\rm cl, 0}}{10^5 \, \msun}\right)  \left( \frac{\ln \Lambda_{\rm c}}{10} \right)\,,
    \label{eq::mvms_simplified}
\end{equation}
with $ m_{\rm seed}$ and $f_{\rm c}$ being the initial mass of the stellar seed and the effective fraction of colliding binaries in the cluster.\\
In general, Eq. \eqref{eq::mvms_simplified} neglects the impact of stellar evolution processes which can affect both the amount of mass effectively accreted onto the growing VMS and the effect of metallicity on mass loss. For instance, strong stellar winds, typical of stellar populations with a metallicity $Z\gtrsim 0.001$, can effectively quench the VMS growth (e.g. \citealt{sevn_spera_2017,iorio_2023}). 
Nonetheless, mass loss is expected to be nearly negligible for metal-poor stellar populations (see for instance Figure 2 of \citealt{mapelli_2018}).
The final mass of the IMBH seeded by the VMS is also highly uncertain, although recent stellar evolution tracks for stars with masses up to $500-2000 \, \msun$ suggest that a relatively small fraction (< 0.2 - 0.3) of the VMS mass is lost during the collapse to an IMBH if the metallicity is $Z< 0.001$ \citep[see Fig. 4 in][]{costa_2025}. Note also that the ultimate fate of a star resulting from multiple stellar collisions may differ from that of a similar star born in isolation due to the different chemical stratification of the collided star and the additional rotation it may acquire during the collisions \citep{glebbeek_2009,costa_stellarcoll_2022, ballone_2023, vergara_2025}.\\

\subsubsection{Mild seeding scenario: stellar collisions in dense star clusters}
\label{subsubsec::seedingindense}
While the runaway scenario requires extreme conditions, in clusters with longer core collapse timescales numerical simulations have shown that several stellar collisions, often triggered by strong dynamical encounters among the population of primordial binaries, can lead to the rapid build-up of IMBH seeds with masses in the range $10 ^2$ up to a few $10^3\,\msun$
\citep[e.g.][]{mapelli_2016, kremer_2020, dicarlo_2021, arcasedda_isolated_2023, frost_rantala_2024, rantala2025rapidchannelcollisionalformation}.
Hence, we consider an additional scenario in which we assume a 20 $\%$ probability that a cluster denser than $\rho_{\rm cl, 0} > 10^5\,\rm \msun\, pc^{-3}$ and with metallicity $Z < 0.001$ forms a VMS, as recently suggested by \cite{dragonII_II_2023} and  \cite{frost_rantala_2024, rantala2025rapidchannelcollisionalformation}. 
Under this scenario, we assign the mass of the VMS randomly between $ m_{\rm VMS, min}$ and $m_{\rm  VMS, max}$, defined as:
    \begin{align}
     m_{\rm VMS, min} &= \min( 150,\, 10^{-0.23} \times (M_{\rm cl, 0} / \msun)^{0.53})\, \msun\,, \\
      m_{\rm VMS, max} &= \min( 2 \times 10^4,\,0.02 \times (M_{\rm cl, 0}/\msun))\, \msun\,.
     \label{eq::mvms_bifrost}
    \end{align} 
The lower bound on the minimum VMS mass matches the values in Table 1 of \cite{frost_rantala_2024} which are derived from the cluster mass – maximum stellar mass relation in \cite{yan_2023}. The upper bound on the maximum VMS mass at $2\times 10^4\,\msun$ serves as a conservative upper bound, given the uncertainties in mass loss and partial disruption of the star during each collision.\\

\subsubsection{IMBH assembly via hierarchical binary black hole mergers}
\label{subsubsec::hierarchical}
Aside from stellar collisions, an IMBH can also grow in a cluster via hierarchical BBH mergers, as summarized below.

Massive objects in a stellar population 
 drift to the center of the parent cluster over a segregation timescale (Eq. \eqref{eq::tseg}), with subsequent three-body \citep{1995MNRAS.272..605L} and binary - single interactions \citep{1993ApJ...403..271G}, promoting the formation of hard binaries \citep{heggie_binary_1975}. 
Though these interactions get progressively rarer, they also become more violent and can impart a recoil to the BBH that may exceed the cluster escape velocity, leading to the premature ejection of the binary. 
However, if the hardening rate due to GW emission becomes dominant over the interaction rate before ejection, the BBH will merge inside the cluster. In such a case, the remnant is either expelled owing to the post-merger relativistic kick, or remains in the cluster possibly participating in another merger event.
In either case, the BBH will undergo the merger only if the sum of cluster formation time, the time needed for the binary to form and harden, and the final merging time does not exceed a Hubble time. 
A BH typically needs to undergo a handful of mergers to reach the IMBH regime, an outcome possible only, in general, if the cluster's escape velocity is sufficiently high to prevent its ejection. 

\begin{table*}[t!]
\centering
\setlength{\tabcolsep}{8pt}
\renewcommand{\arraystretch}{1.5}
\begin{tabular}{lp{5cm}p{3cm}p{4cm}}
\toprule
Model & IMBH formation & Seeding conditions & Formation histories \\ 
\midrule
A & Runaway stellar collisions and/or hierarchical BBH mergers & $t_{\rm cc} < 5 \rm\,Myr$, \newline $Z < 0.001$ & Uniform for $z \in [2, 8]$ \\
\\
A' & Same as Model A & Same as Model A & Normal dist. around $z \sim 3.2$, follows galactic SFR \\[0.2cm]
\midrule
B & Repeated stellar collisions and/or hierarchical BBH mergers & $\rho_{\rm cl, 0} > 10^5\, \rm \msun\, pc^{-3}$, $Z < 0.001$, $P_{\rm seed} = 0.2$ & Same as Model A \\[0.2cm]

\bottomrule
\end{tabular}
\caption{Models considered in this work and their fundamental differences in IMBH seeding conditions and initial redshift distributions of \globular and \nuclear.}
\label{tab:imbh-models}
\end{table*}


\subsection{Models descriptions}
\label{subsec::models}
We explore three models based on different prescriptions for the IMBH seeding or the formation history of star clusters. The models are summarized in the following and in Table \ref{tab:imbh-models}:
\begin{itemize}
    \item \textit{Model A}: All clusters with a core collapse time shorter than $5\,\rm Myr$ and metallicity $Z < 0.001$ are assumed to trigger the formation of a VMS via runaway stellar collisions, followed by the seeding of an IMBH, as described in Section \ref{subsubsec::runaway}. We also invoke for the seeding a minimum cluster mass of $2\times 10^4 \, \msun$ in order to have, on average, at least one seed massive star in the cluster (see Appendix \ref{app::depletion}). The VMS mass ($m_{\rm VMS}$) is computed according to Eq. \eqref{eq::mvms_simplified} assuming an initial star mass $m_{\rm seed} = 150\,\msun$ and an effective fraction of colliding binaries $f_{\rm c} = 0.2$. The resulting IMBH is assumed to retain 80 $\%$ of the VMS mass, consistently with the findings of \cite{costa_2025} for these metallicities. We then follow the dynamical processes that may determine the pairing and merger of the IMBH with other stellar-mass BHs. 
    If core collapse happens after $5\,\rm Myr$, the runaway process is suppressed, and 
    the IMBH formation can occur only through hierarchical BBH mergers (Section \ref{subsubsec::hierarchical}).\\
    In this model, we assume \young to follow the SFR from MF17, while for both \globular and \nuclear we adopt the redshift distribution of KR13. The latter assumption supports the scenario in which \nuclear share their formation history with \globular, as might occur if \nuclear form through the (possibly rapid) merger of \globular migrated towards the galactic center via dynamical friction, a scenario commonly referred to as the "migration" scenario \citep[and references therein]{tremaine1975_ncform,capuzzo1993_ncform, neumayer_review2020}.
    \\
    \item \textit{Model A'}: This model adopts the same assumptions of Model A, except for the formation histories of \globular and \nuclear. As with \young, we assume \nuclear to follow the SFR from MF17. This choice reflects a scenario in which \nuclear form directly through star formation and gas infall occurring near the galactic center, a process commonly referred to as the "in-situ" formation scenario \citep{loose1982}.
    In addition, we sample the \globular formation redshift according to a Gaussian distribution with mean redshift $\mu_z=3.2$ and $\sigma_z =0.2$ \citep{mapelli_2022}. This choice reflects theoretical models aimed at tracking the formation of Galactic \globular (see e.g. EB18).
    \\
    \item \textit{Model B}: This model is based on the same set of assumptions of Model A regarding the cluster formation history, but 
    assumes a mild seeding scenario in which stellar collisions can lead to the formation of a VMS with a $20 \%$ probability in clusters that are both metal-poor, $Z< 0.001$, and very dense, $\rho_{\rm cl, 0} > 10^5\, \msun\,\rm pc^{-3}$ (see also Section \ref{subsubsec::seedingindense}). As for the runaway seeding, we fix a minimum cluster mass ($5 \times 10^3  \,\msun$) to ensure the presence of massive stellar seeds for the collisions (see Appendix \ref{app::depletion}). 
    The mass of the VMS is sampled as in Eq. \ref{eq::mvms_bifrost} while we assume the IMBH to be $80 \% $ the mass of its progenitor. As in Model A simulation, if the cluster does not meet the seeding conditions we simply follow the dynamical evolution of a stellar-mass BH which may grow into an IMBH via hierarchical BBH mergers.\\ 
\end{itemize}

\section{Results}
\label{sec:results}
Hereafter, we refer to IMBHs formed from a stellar BH that underwent a chain of BBH mergers as "hierarchical IMBHs", while we refer to IMBHs formed through direct collapse of a stellar collision product as "IMBH seeds" or "seeded IMBHs". We point out that IMBH seeds can also grow by further merging with stellar BHs.

\begin{figure*}[t]
    \centering
    \includegraphics[width=1.9\columnwidth]{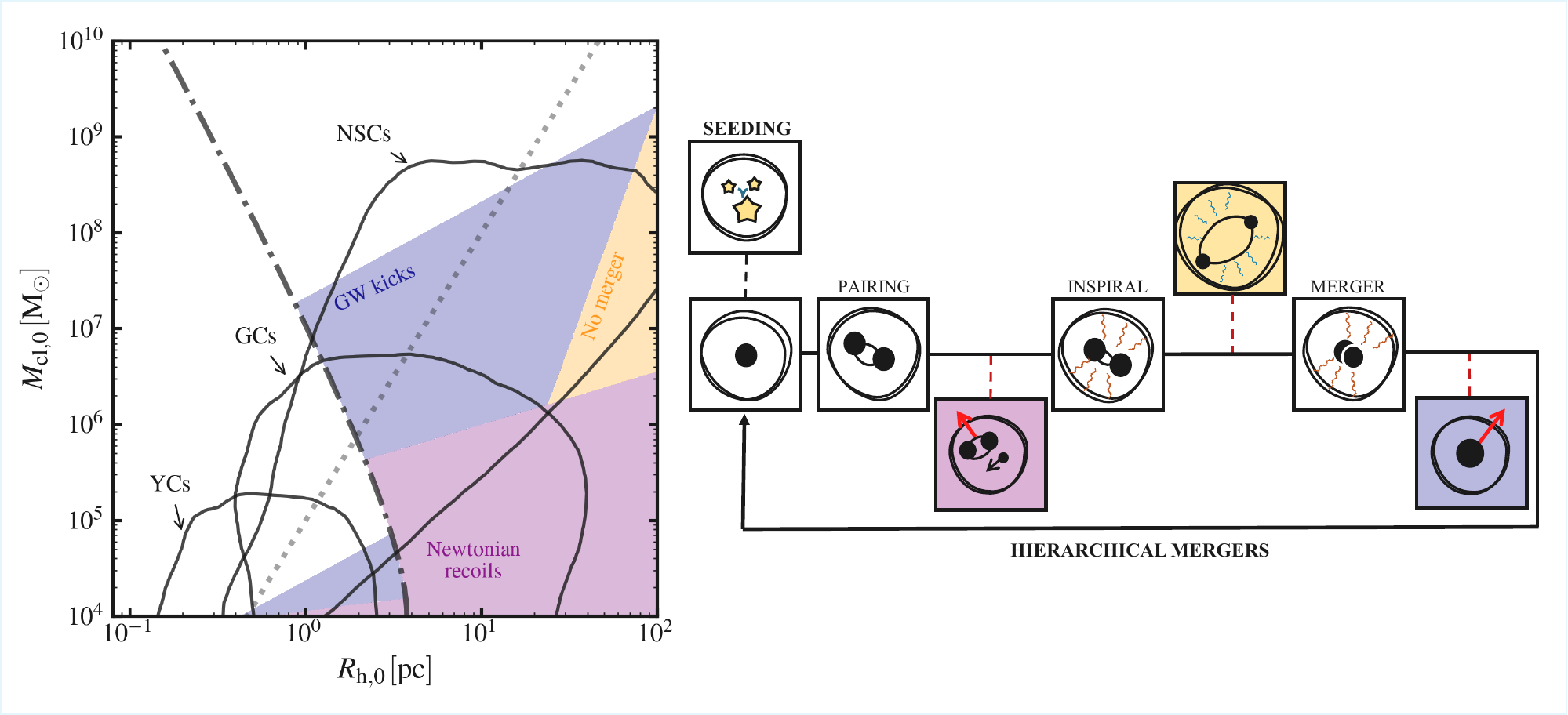}
    \caption{
        \small
        \textit{Left panel:} Initial cluster mass and half-mass radius for \young (bottom left area), \globular (central area),  and \nuclear (upper right area). The clusters distribution are cut at the 99$\%$ contour.
        Clusters with 
        $t_{\rm cc} < 5\,\rm Myr$ lie below the dot-dashed black line, those with central density $\rho_{\rm cl, 0} > 10^5\,\msun \rm \, pc^{-3}$ above the shaded dotted gray line. We also identify regions of the parameter space in which (i) a BBH binary is ejected before merging due to strong Newtonian recoils ("Newtonian recoils" region, in pink), (ii) the merger remnant is tipically kicked out of the cluster due to its relativistic kick ("GW kicks" region, in purple), (iii) the BBH merger time exceeds a Hubble time ("No merger" region, in yellow).\\
        \textit{Right panel:} Schematic overview of a BH dynamical growth in a star cluster. The colored boxes refer to the regions highlighted in the left plot.}
    
    \label{fig::MRdiag}
\end{figure*}

\subsection{IMBH formation efficiency}
\label{subsec::mrdiagram}

In Figure \ref{fig::MRdiag} left panel, we present the input distribution of initial masses and half-mass radii of \young, \globular and \nuclear used in our simulations.

The runaway stellar collisions seeding, which we assume in Model A and A', comprises all the environments falling on the left side of the dashed-dotted contour at $t_{\rm cc} = 5 \,\rm  Myr$.
As shown in the plot, this seeding mechanism typically favors \young as well as very light \globular and \nuclear. Approximately $11\%$ of \young, $39\%$ of \globular, and $20\%$ of \nuclear in \bpop fall within this runaway seeding region.

Conversely, the mild collision scenario, which we assume in Model B, selects the environments which fall above the gray dotted density contour at (half-mass) density $\rho_{\rm cl, 0} = 10^5\,\msun\,\rm pc^{-3}$. This seeding condition generally favors \nuclear, which present the densest environments among the different cluster families. In this scenario we find that $\sim 19 \%$, $9 \%$ and $32 \%$ of \young, \globular and \nuclear mass-radius distributions fall in the portion of the parameter space where $\rho_{\rm cl, 0} > 10^5\,\msun\,\rm pc^{-3}$. Assuming an additional 20 \% seeding probability for these clusters (see more in Section \ref{subsubsec::seedingindense}) leads to $\sim 4 \%$, $2 \%$ and $6 \%$ of \young, \globular and \nuclear being potential IMBH seeds hosts in Model B.

Note, however, that the actual fraction of clusters producing IMBH seeds in both models depends critically on the metallicity, as we assume that the VMS progenitors are assembled only in clusters with $Z < 0.001$. The impact of metallicity is strictly linked to the clusters formation history and can significantly decrease the effective fraction of seeds hosts as we discuss in more details in Section~\ref{sec::demographics}.

The efficiency of hierarchical BBH mergers in assembling an IMBH is more complicated to define since many processes can actually lead to the ejection of the original BH during its growth. As a reference, in Figure \ref{fig::MRdiag} we present, in different colors, the regions of the parameter space where a BBH (i) is ejected prior to merger due to a strong Newtonian recoil, (ii) merges and produces a BH which is then ejected by its relativistic kick, or (iii) the BBH is retained in the cluster but fails to merge within a Hubble time 
(see Appendix \ref{app::theory} for further details). The first two regions are drawn comparing the magnitude of the recoils and kicks to the initial escape velocity of the host clusters. To evaluate the GW kicks we average over all possible spin configurations assuming them to be isotropically distributed in space, and sampling the spins magnitudes from our fiducial Maxwellian distribution (see more in Section \ref{sec::methods}). The last region is evaluated computing the average delay time between the BBH formation and merger in the cluster, and comparing it to a Hubble time.
We adopt a $30\,\msun-30\,\msun$ BBH as our reference system in all clusters, except when the cluster lies in the regime of runaway stellar collisions (i.e. $t_{\rm cc}<5\,\mathrm{Myr}$). In that situation, we instead consider an IMBH-BH binary, with the primary mass set to $m_1 = 0.8\,m_{\rm VMS}$ (Eq.~\eqref{eq::mvms_simplified}) and the secondary mass to $m_2 = 30\,\msun$. For the stellar-mass BBH case, only the most massive \nuclear lie outside the regions where the binary is either ejected or prevented from merging. In contrast, when an IMBH-BH binary forms from a seeded IMBH, the larger inertia of the system significantly reduces the probability of ejection, making retention within the cluster more likely 
even for \young\ and \globular.
This picture becomes, however, far more complicated when accounting for the temporal evolution of the host cluster which can strongly impact the IMBH formation and growth through the hierarchical channel.
For instance, the cluster mass loss and expansion will drive clusters to wander downward and leftward in the $M_{\rm cl, 0} - R_{\rm h, 0}$ plane shown in Figure \ref{fig::MRdiag}, reducing the BH interaction rate and increasing the probability of BH ejection. This is especially the case of clusters with relaxation times comparable to the dynamical timescales over which the BBHs evolve. We will explore all these aspects in more detail in the next section.
Finally, in Figure \ref{fig::MRdiag} we also provide a simple sketch of the processes contributing to the seeding and growth of an IMBH in a star cluster as well as the ejection mechanisms mentioned above.

\subsection{Demographics}
\label{sec::demographics}
We find that only a small fraction of the $10^8$ BH realizations produces an IMBH: about $\sim 0.6\%$ in Model A, $\sim 0.4\%$ in Model A', and $\sim 0.1\%$ in Model B. In all cases, the overwhelming majority of IMBHs originates from stellar collision seeds. Specifically, these IMBHs represent $\sim 97 - 98\%$ of the whole IMBH population in Models A and A', and $\sim 88\%$ in Model B. 
Hence, in all our simulations stellar collisions appear to be the primary drivers of IMBH formation in star clusters.
The overall IMBH fractions remain modest mainly due to the imposed constraint on the metallicity threshold which we require for the formation of a VMS via stellar collisions. It is also important to note that, in our simulations, \young, \globular, and \nuclear are assumed to be equally abundant. Therefore, these fractions do not reflect the true relative proportions of different cluster families in the Universe, which remain highly uncertain. For reference, the Milky Way hosts one NSC, $\sim 200$ \globular\ \citep{harris_1996, minniti_2017}, and possibly up to 
$\sim 10^3$ massive \young\ clusters \citep{larsen_2009}.

To quantify the IMBH formation efficiency within each cluster family, we report in Table~\ref{tab:cluster-distribution2} the final IMBH populations, distinguishing between their formation channel (hierarchical or seeded) and the host cluster type (\young, \globular, \nuclear). The percentages shown for each mechanism are computed relative to the total number of clusters belonging to the corresponding family. From Table~\ref{tab:cluster-distribution2} it is clear that seeding via stellar collisions is consistently more efficient than hierarchical growth, by at least one order of magnitude, regardless of the cluster type. 
The environments in which this mechanism is more frequent, however, depend strongly on both the seeding conditions (Model A vs. Model B) and the assumed cluster formation histories (Model A vs. Model A'). 
For instance, in Model A, \globular are more likely to be hosting a seed thanks to their shorter core collapse times, whereas in Model B the seeding is restricted to very dense clusters, hence IMBH seeds are formed more frequently in \nuclear. 
A direct comparison between Models A and A' further illustrates the impact of formation history. In Model A', only about $0.05\%$ of \globular\ produce an IMBH seed, compared to $\sim 0.8\%$ of \nuclear. This trend is reversed in Model A, where \globular\ are nearly twice as likely as \nuclear\ to host a seed. The difference arises because, in Model A', \globular\ statistically form at lower redshifts than in Model A, and thus tend to have higher metallicities, often exceeding the threshold for runaway or mild stellar collisions. While \nuclear\ also tend to form later in Model A', their redshift distribution extends up to $z \gtrsim 10$, broader than that of \globular. This wider redshift range partly offsets the metallicity constraint, allowing a non-negligible fraction of \nuclear\ to still meet the seeding conditions.

Across all three models, hierarchical IMBH formation is efficient only in \nuclear, while \young\ and \globular\ produce virtually no IMBHs through this channel (see Table~\ref{tab:cluster-distribution2}). This is because only \nuclear\ reach escape velocities high enough to retain stellar-mass BHs against the typical dynamical recoils and relativistic kicks they experience during their growth (see Section~\ref{subsec::finalfate}). 

\begin{table}[h]
  \centering
  \renewcommand{\arraystretch}{1.3}
  \setlength{\tabcolsep}{6pt}
  \small
  \begin{tabular}{lccc}
  \multicolumn{4}{c}{\textit{Model A} } \\
\toprule
Formation & \young [$\%$] & \globular [$\%$] & \nuclear [$\%$] \\
\midrule
Seeded       & 0.38     & 0.94     & 0.51 \\
Hierarchical & $<$ 0.01 & $<$ 0.01 & 0.04 \\
\midrule
All          & 0.38     & 0.94     & 0.55 \\
\bottomrule
    \\
  \multicolumn{4}{c}{\textit{Model A'} } \\
\toprule
Formation & \young [$\%$] & \globular [$\%$] & \nuclear [$\%$] \\
\midrule
Seeded       & 0.38     & 0.05     & 0.79 \\
Hierarchical & $<$ 0.01 & $<$ 0.01 & 0.03 \\
\midrule
All          & 0.38     & 0.05     & 0.82 \\
\bottomrule
\\
  \multicolumn{4}{c}{\textit{Model B} } \\
\toprule
Formation & \young [$\%$] & \globular [$\%$] & \nuclear [$\%$] \\
\midrule
Seeded       & 0.04     & 0.04     & 0.16 \\
Hierarchical & $<$ 0.01 & $<$ 0.01 & 0.04 \\
\midrule
All          & 0.04     & 0.04     & 0.20 \\
\bottomrule
\end{tabular}
  \caption{\small Fractions of IMBHs per formation scenario (Seeded or Hierarchical) and cluster host (\young, \globular, \nuclear). The percentages are normalized to the total number of clusters simulated per each cluster type.}
  \label{tab:cluster-distribution2}
\end{table}
\subsection{Mergers and IMRIs}
The percentages in Table \ref{tab:cluster-distribution2} are evaluated on the overall population of simulated BHs ($\sim$ 1/3 of $10^8$ BHs per cluster family). However, only a subset of these BHs actually undergoes at least one merger, specifically those for which the overall merger times are smaller than a Hubble time (see Appendix \ref{app::theory} for details). In Table \ref{tab:imbh_mergers} we present the different merger efficiencies ($\epsilon_{\rm BBHs}$) we find for \young, \globular and \nuclear. These efficiencies represent the fraction of BHs that experience at least one merger out of the total simulated BH population. Their values are the same for Model A and B but change slightly in Model A' for \globular and \nuclear. This indicates that the choice of seeding prescription has only a minor impact on the overall number of merging BHs. In contrast, differences in the assumed formation histories for \globular and \nuclear play a more significant role.
In Table \ref{tab:imbh_mergers} we also report the fraction of merging BHs which ultimately result in the merger between an IMBH and a stellar-mass BH, also known as intermediate mass-ratio inspirals (IMRIs)\footnote{For hierarchical BBH mergers this means that the BH undergoes a sufficient number of mergers to exceed $100 \, \msun$, whereas for IMBH seeds an IMRI forms as soon as the seed pairs up with a stellar-mass BH and merges within a Hubble time.}.
These systems are particularly interesting since they can represent potential GW sources detectable by the next generation detectors in different frequency bands, from mHz, like LISA \citep{2024arXiv240207571C}, to deci-Hz \citep{2020CQGra..37u5011A}, like the Lunar Gravitational-Wave Antenna \citep{2025JCAP...01..108A}, to Hz and beyond, like the Einstein Telescope \citep{2025arXiv250312263A}. 
We evaluate the IMRIs fractions per each cluster family, along with the median number of mergers ($\langle N_{\rm mer}\rangle$) the progenitor IMBH or BH undergoes in that specific environment, and the sub-fractions of IMRIs in which the primary is heavier than $150 \, \msun$ or $10^3 \, \msun$, respectively.

As expected, IMRIs typically arise from IMBH seeds, with relative abundances up to two orders of magnitudes larger than the ones observed for the hierarchical channel depending on the environment and on the model considered. More specifically, the hierarchical channel is extremely inefficient compared to the seeding one in \young while the two channels produce comparable amounts of IMRIs in \nuclear. Significant differences in the IMRI fractions also emerge between the two 
seeding models (Model A and Model B), suggesting that future detections of 
these systems could serve as a valuable diagnostic for discriminating among 
different seeding mechanisms. The actual detectability of these sources, however, will naturally depend on the sensitivity of specific interferometers; evaluating event rates for different instruments will be our next step to translate these findings into observational forecasts.

Seeded IMBHs often undergo many more mergers (on average) than hierarchical IMBHs. Indeed, IMBH seeds are much heavier at birth than the underlying stellar-mass BH populations in the host clusters and are less prone to ejection (see more in Section \ref{subsec::finalfate}), hence sustaining longer merger chains. This difference across the seeded and hierarchical channel is particularly clear in Model B, in which IMBHs seeds form in very dense environments which favor the retention and fast mergers of the seed with other BHs. In \nuclear half of the IMBH seeds can exceed $\sim 50$ ($\sim 240$) mergers in Model A (B). 
In contrast, in \young IMBHs seeds typically undergo only one merger before being expelled from their host. Similarly, hierarchical IMBHs forming in \young and \globular typically emerge from a single BBH merger of relatively heavy stellar-mass BHs. Most of these IMBHs present masses at the lowest end of the IMBH mass spectrum ($\sim 100 \, \msun$), hence no IMRI is found with primary mass above $150 \,\msun$ in these environments ($f_{> 150 \,\msun} = 0$). 
High-mass IMRIs ($> 10^3 \, \msun$) occur at a rate of approximately one per $10^2$--$10^4$ BBH mergers in our models. The production of such heavy binaries is especially interesting for \young\ and \globular, where, as we have previously mentioned, the pure hierarchical formation channel is otherwise highly inefficient.

\begin{table}[htbp]
  \centering
  \small
  \renewcommand{\arraystretch}{1.3}
  \setlength{\tabcolsep}{4pt}
  \begin{tabular}{@{}lcccccc@{}}
  \multicolumn{7}{c}{\textit{Model A}} \\
\toprule
Clu. & $\epsilon_{\rm BBHs}$ & Form. & $f_{\rm IMRIs}$ & $\langle N_{\text{mer}}\rangle$ & $f_{>150 \, \msun}$ & $f_{>10^3 \, \msun}$ \\
\midrule
\multirow{2}{*}{\young} 
& \multirow{2}{*}{$1 \times 10^{-2}$} & S  & $3 \times 10^{-1}$ & 1   & $3 \times 10^{-1}$ & $8 \times 10^{-3}$ \\
&                                    & H  & $1 \times 10^{-3}$ & 1   & -- & -- \\
\midrule
\multirow{2}{*}{\globular} 
& \multirow{2}{*}{$1 \times 10^{-1}$} & S  & $9 \times 10^{-2}$ & 8   & $9 \times 10^{-2}$ & $5 \times 10^{-2}$ \\
&                                    & H  & $2 \times 10^{-4}$ & 1   & -- & -- \\
\midrule
\multirow{2}{*}{\nuclear} 
& \multirow{2}{*}{$3 \times 10^{-1}$} & S  & $2 \times 10^{-2}$ & 50  & $2 \times 10^{-2}$ & $1 \times 10^{-2}$ \\
&                                    & H  & $1 \times 10^{-3}$ & 4   & $3 \times 10^{-4}$ & $2 \times 10^{-4}$ \\
\bottomrule
\\
\multicolumn{7}{c}{\textit{Model A'}} \\
\toprule
Clu. & $\epsilon_{\rm BBHs}$ & Form. & $f_{\rm IMRIs}$ & $\langle N_{\text{mer}}\rangle$ & $f_{>150\, \msun}$ & $f_{>10^3\, \msun}$ \\
\midrule
\multirow{2}{*}{\young} 
& \multirow{2}{*}{$1 \times 10^{-2}$} & S  & $3 \times 10^{-1}$ & 1   & $3 \times 10^{-1}$ & $9 \times 10^{-3}$ \\
&                                    & H  & $2 \times 10^{-3}$ & 1   & -- & -- \\
\midrule
\multirow{2}{*}{\globular} 
& \multirow{2}{*}{$6 \times 10^{-2}$} & S  & $8 \times 10^{-3}$ & 8   & $8 \times 10^{-3}$ & $5 \times 10^{-3}$ \\
&                                    & H  & $2 \times 10^{-5}$ & 1   & -- & -- \\
\midrule
\multirow{2}{*}{\nuclear} 
& \multirow{2}{*}{$2 \times 10^{-1}$} & S  & $4 \times 10^{-2}$ & 50  & $4 \times 10^{-2}$ & $3 \times 10^{-2}$ \\
&                                    & H  & $1 \times 10^{-3}$ & 2   & $2 \times 10^{-4}$ & $1 \times 10^{-4}$ \\
\bottomrule
\\
\multicolumn{7}{c}{\textit{Model B}} \\
\toprule
Clu. & $\epsilon_{\rm BBHs}$ & Form. & $f_{\rm IMRIs}$ & $\langle N_{\text{mer}}\rangle$ & $f_{>150\, \msun}$ & $f_{>10^3\, \msun}$ \\
\midrule
\multirow{2}{*}{\young} 
& \multirow{2}{*}{$1 \times 10^{-2}$} & S  & $4 \times 10^{-2}$ & 1   & $4 \times 10^{-2}$ & $2 \times 10^{-3}$ \\
&                                    & H  & $3 \times 10^{-3}$ & 1   & -- & -- \\
\midrule
\multirow{2}{*}{\globular} 
& \multirow{2}{*}{$1 \times 10^{-1}$} & S  & $5 \times 10^{-3}$ & 75  & $5 \times 10^{-3}$ & $4 \times 10^{-3}$ \\
&                                    & H  & $2 \times 10^{-4}$ & 1   & $6 \times 10^{-7}$ & -- \\
\midrule
\multirow{2}{*}{\nuclear} 
& \multirow{2}{*}{$3 \times 10^{-1}$} & S  & $6 \times 10^{-3}$ & 243 & $6 \times 10^{-3}$ & $5 \times 10^{-3}$ \\
&                                    & H  & $1 \times 10^{-3}$ & 4   & $3 \times 10^{-4}$ & $2 \times 10^{-4}$ \\
\bottomrule
\end{tabular}
\caption{\small Summary of IMBH and BBH formation per cluster type - i.e. \young, \globular or \nuclear~- and formation channel - Hierarchical (H) or Seeded (S). Columns 2, 4, 5, 6 and 7 show the BBH merger efficiency $\epsilon_{\rm BBHs}$, the IMRI fraction $f_{\rm IMRIs}$, median number of mergers $\langle N_{\rm mer}\rangle$, and the fractions of IMBHs above $150 \, \msun$ and $10^3 \, \msun$, respectively.}
\label{tab:imbh_mergers}
\end{table}

\subsection{IMBHs at low redshifts: models and observations}

\subsubsection{Retention efficiency and ejection mechanisms}
\label{subsec::finalfate}
At redshift $z=0$, only around 33$\%$ (Model A), 39$\%$ (Model A'), and 61$\%$ (Model B) of all IMBHs forming in the simulations are retained in their clusters. 
In general, the IMBHs in our simulations can be ejected by means of three different processes: (i) Newtonian recoils, which expel an IMBH-BH binary as a result of strong binary-single interaction with the other BHs in the cluster, (ii) relativistic kicks, which eject an IMBH remnant after the binary merged, and (iii) the complete evaporation of the IMBH host cluster. The latter process mainly affects light clusters with short relaxation times which completely disrupt within a Hubble time, releasing their IMBHs in the field (see also Section \ref{subsec::lowredcandidates}).

In Table~\ref{tab:imbhs-final-fate}, we summarize the final fate of IMBHs in our simulations, categorized by formation mechanism, cluster family, and ejection process. We also report the fraction of IMBHs retained at $z = 0$, with all percentages normalized to the total number of IMBHs formed in each cluster type and formation scenario. We focus on the comparison between Model A and B, as Model A' exhibits trends similar to Model A. Indeed, while Model A and A' differ in the total number of IMBH seeds due to the different choice of initial metallicity (redshift) distributions (Section~\ref{sec::demographics}), the relative fractions of IMBHs that are ejected or retained remain essentially unchanged. This is because these fractions are primarily set by the initial BH and IMBH seed properties and by the escape velocities and dynamical evolution of their host clusters, which are identical in both models. In addition, for hierarchical IMBHs we only refer to \nuclear, since the fractions of IMBHs formed in \young and \globular through this channel are negligible (see Section \ref{sec::demographics}).

Model B exhibits a substantially higher fraction of IMBH seeds retained at $z = 0$ compared to Model A. This arises from the higher-density threshold imposed for IMBH seeding in Model B, which preferentially selects host clusters with larger escape velocities than those in Model A.
In both models, IMBHs are primarily ejected through post-merger relativistic kicks. This is particularly evident for hierarchical IMBHs, which are retained at $z = 0$ in less than 15$\%$ of the times.
In addition to dynamical ejection mechanisms, such as kicks or recoils, we find that the premature evaporation of \young quenches the growth of IMBH seeds in $11 - 17\%$ of cases,  depending on the model. 
In Section \ref{disc::wandering}, we discuss possible observational implications for this population of orphan IMBHs.

\begin{figure*}[htb!]
    \centering
    \includegraphics[width=1.8\columnwidth]{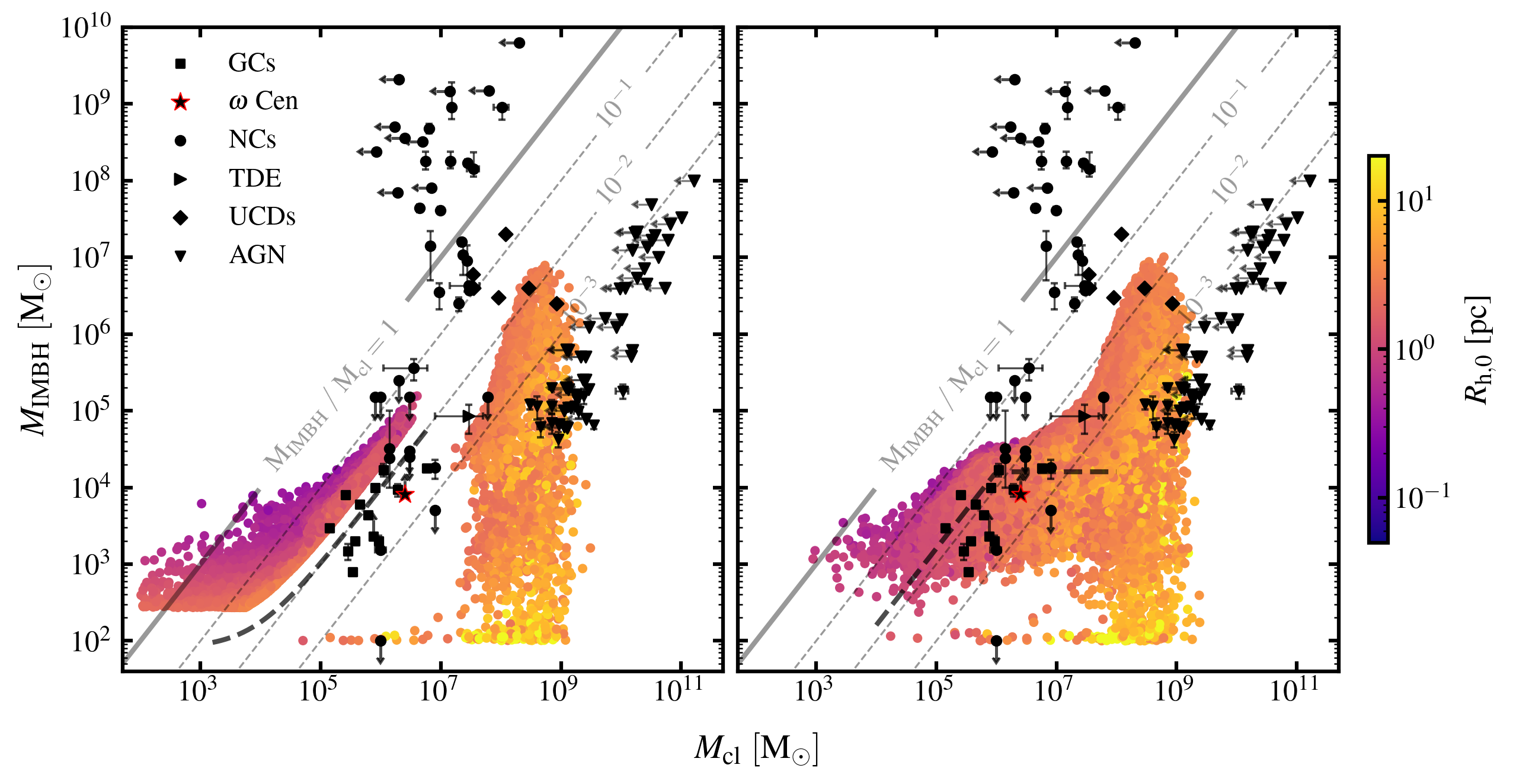}
    \caption{\small Clusters hosting an IMBH at $z = 0$ in Model A (left panel) and Model B (right panel) compared to different classes of IMBH host candidates observations. 
    Note that the masses reported for low-mass AGN refer to the host galaxy mass (up to $\sim$ 100 times larger than the mass of the galactic nuclei). 
    The gray dashed line shows the theoretical prediction for the initial mass of an IMBH seeded via stellar collisions (Eq. \eqref{eq::mvms_simplified} in the left panel and Eq. \eqref{eq::mvms_bifrost} in the right panel).}
    \label{fig::allIMBHs}
\end{figure*}

\begin{table}[h]
\centering
\setlength{\tabcolsep}{3pt}
\renewcommand{\arraystretch}{1.3}
\small
\begin{tabular}{@{}llcccc@{}}
\multicolumn{6}{c}{\textit{Model A}} \\
\toprule
Form. & Clu. & N. Rec. [\%] & GW Kick [\%] & Evap. [\%] & Ret. [\%] \\
\midrule
\multirow{3}{*}{S}
 & \young    & 14.2 & 74.0 & 11.8 & -- \\
 & \globular & 4.1  & 61.5 & 0.6  & 33.8 \\
 & \nuclear  & 1.8  & 39.5 & 0.5  & 58.2 \\
\midrule
H & \nuclear & 0.8  & 85.6 & -- & 13.6 \\
\midrule
\multicolumn{6}{c}{\textit{Model B}} \\
\toprule
Form. & Clu. & N. Rec. [\%] & GW Kick [\%] & Evap. [\%] & Ret. [\%] \\
\midrule
\multirow{3}{*}{S}
 & \young    & 21.0 & 61.6 & 17.4 & -- \\
 & \globular & 0.8  & 34.3 & 0.5  & 64.4 \\
 & \nuclear  & 0.3  & 11.0 & 0.1  & 88.5 \\
\midrule
H & \nuclear & 1.0  & 84.7 & -- & 14.4 \\
\bottomrule
\end{tabular}
\caption{
\small
Percentage of IMBHs subdivided by formation channel (Column 1) and cluster type (Column 2) which are expelled at $z > 0$ as a result of (i) a strong Newtonian recoil (Column 3), (ii) a post-merger relativistic kick (Column 4), or (iii) the evaporation of their host cluster (Column 5). We also present the fraction of IMBHs which are retained within their hosts (Column 6).}
\label{tab:imbhs-final-fate}
\end{table}

\subsubsection{Comparison to low-redshift IMBH candidates}
\label{subsec::lowredcandidates}
In Figure \ref{fig::allIMBHs}, we compare the IMBH population retained at $z= 0$ in our simulations with 
several observations of IMBH candidates in local \globular \citep{lutz2013_gcs, kiziltan_2017, haberle_omegacen_2024}, in extragalactic clusters from potential tidal distruption events \citep{tde_lin_2018}, in galactic nuclei (\citealt{ncs_graham_2009, ncs_neumayer_2012, ncs_kormendy_2013}), 
in dwarf galaxies, ultra compact dwarf galaxies (UCDs) and low-mass active galactic nuclei (AGN) \citep{reines_2015,chilingarian_18_IMBHsAGNs, nguyen_18_ucds}.
In the following, 
we restrict the analysis to Model A and Model B, as Model A and Model A' produce results broadly consistent with each other. Model A is shown in the left panel of Figure \ref{fig::allIMBHs}, while Model B is displayed in the right panel.

In Model A, IMBH seeds can form only in clusters with a core collapse time $t_{\rm cc}\lesssim 5\,\rm Myr$, a condition that translates into  an upper limit on the clusters relaxation times and, consequently, their lifetime.
This effect is evident in Figure \ref{fig::allIMBHs} for \young and low-mass \globular, which in our simulations either evaporate before $z=0$ (see also Table \ref{tab:imbhs-final-fate}) or undergo substantial mass loss, retaining IMBHs with typical final mass ratios $M_{\rm IMBH}/ M_{\rm cl} \gtrsim 0.1$. 
Conversely, hierarchical IMBHs are retained only in clusters with final masses $\gtrsim 10^7\,\msun$ and exhibit typical mass ratios $M_{\rm IMBH}/M_{\rm cl} \lesssim 10^{-2}$, filling a region of the cluster mass - IMBH mass plane partly populated by observed \nuclear. The growth proceeds uninterrupted in the densest clusters with final masses $> 10^8 \, \msun$, generally leading to the exhaustion of the BH reservoir. These clusters show notable similarities with more complex stellar systems, such as UCDs and light low-mass AGN.
In Model A we observe a distinct "gap" around $M_{\rm cl}\sim 10^7 ,\msun$ between the seeded and hierarchical populations, which lie to the left and right of the gap, respectively. The gap emerges from the combined effect of the seeding conditions and the inefficiency of hierarchical BBH mergers in clusters falling in that mass range. For completeness, in Appendix \ref{app:gap_position} we also discuss how the gap may be affected assuming \globular to be more massive and compact at birth.
The IMBH mass -- host mass relation changes significantly in Model B. In this model the seeded and hierarchical retained populations fill homogeneously the cluster mass range $\sim 10^3-10^9\,\msun$ and clearly overlap with the masses of putative IMBHs observed in galactic and extragalactic \globular, as well as \nuclear in the local Universe. In both models we observe a tail of light IMBHs ($\sim 100 \, \msun$) in clusters heavier than $\sim 10^5 \, \msun$. These IMBHs are retained as part of IMRIs with merging times exceeding a Hubble time.

The clear difference between the two seeding models at $z=0$ raises several interesting points. First, it highlights the limited role of runaway stellar collisions in favoring the growth of IMBHs heavier than $10^4 - 10^5 \, \msun$. Second, it clearly shows the low efficiency of the hierarchical BBH merger scenario in explaining low-redshift IMBHs observations, except in very massive and dense hosts ($\gtrsim 10 ^7 \, \msun \,  \rm pc^{-3}$). Third, it suggests that a hybrid scenario in which an IMBH seed forms from a few collisions among massive stars (e.g., triggered by binary-single and binary-binary interactions) and possibly further grows by merging with lighter BHs, as in Model B, may be a promising dynamical formation channel for IMBHs in GC-like environments.

\section{Discussion}
\label{sec::discussion} 
In the following, we present some practical applications of our models to the study of IMBHs in \young and \globular. In particular, we analyze the implications of premature IMBH ejection from \young on wandering IMBHs in the Milky Way and other galaxy halos, and investigate the origin of Galactic \globular through potential IMBH signatures. These discussions will form the basis for more detailed studies in forthcoming papers.

\subsection{Young clusters and wandering IMBHs}
\label{disc::wandering}
As discussed in Section \ref{subsec::lowredcandidates}, metal-poor \young with short relaxation time can form IMBHs via stellar collisions. However, many of these IMBHs are likely to be ejected shortly after formation due to Newtonian recoils, post-merger relativistic kicks, or the evaporation of their host clusters. In this section, we briefly explore whether a population of IMBHs formed in Milky Way young and open clusters could remain trapped in the Galactic potential and now wander across the Milky Way.

To explore this scenario and filter out statistical fluctuations, we re-simulate $5 \times  10^7$ BBHs in \young, focusing on the runaway collision scenario (Model A) and the mild collision scenario (Model B). In all models, IMBHs are either ejected during their build-up, or are retained until the cluster evaporates via internal relaxation and tidal disruption operated by the galactic potential.
Upon cluster dispersal, the IMBH is "gently" released in the galactic environment, where it will keep following the original orbit of its former host up to an Hubble time. Clusters formed sufficiently close to the center of the galaxy can also drift inward and bring their IMBHs to the galactic nucleus, a process that possibly impacts both the growth of \nuclear and SMBHs \citep{tremaine1975_ncform, capuzzo1993_ncform, ebisuzaki_2001, pz_2006, antonini_2012, arcasedda_2015, arcaseddagualandris_2018}. \\
Depending on the ejection mechanism, the IMBH is released in the field with a different velocity. IMBHs ejected via GW relativistic kicks receive a median kick of $10^2-10^4\,\rm km/s$, those ejected through dynamical interactions attain velocities of the order of $\sim 10\,\rm km/s$, while for those released by evaporating clusters we assume a velocity $\sim 0\,\rm km/s$. The ejection mechanism also determines whether the IMBH is expelled as a single object (e.g. in the aftermath of a BBH merger) or as a binary (in case of dynamical ejections and evaporated clusters). 
Regardless of the ejection mechanism, IMBH ejection velocities are generally much lower than the typical escape velocity estimated for the present-time Milky Way disk ($\sim 500\,\rm km/s$, see for reference \citealt{gaia_2025}), thus suggesting that expelled IMBHs should remain anchored to the Milky Way gravitational field.

To constrain the number of IMBHs produced in \young and now wandering the Milky Way, we assume that the Galactic potential at the time of ejection resembles its present-day configuration. 
As a reference, we adopt the \texttt{MWPotential2014} model implemented in \texttt{galpy} \citep{galpy_2015}, which combines the contributions of a bulge, a disk, and a dark matter halo.

Observations of Milky Way disk clusters suggest that $\sim 19000$ \young with masses $\gtrsim 10^4\,\msun$ may have formed over $\sim 10$ Gyr \citep{larsen_2009}. Using our catalogs, we generate 5000 realizations of this population, drawing 19000 \young per realization with initial masses above $10^4\,\msun$. Each YC (IMBH) is assigned a random position in the galactic plane, and the number of IMBHs retained is determined by comparing their final recoil or kick to the Galactic escape velocity. Our analysis suggests that $273 \pm 8$ ($26 \pm 2$) wandering IMBHs form in Model A (Model B), among which roughly $\sim 20 - 25 \%$ are in IMBH-BH systems ($48 \pm 4$ for Model A and $8 \pm 1$ for Model B). 
In both models, the majority of these IMBHs have masses around $\sim 350\,\msun$ and ejection velocities $v_{\rm ej} < 10^2$ km/s.

The potential existence of these IMBHs in the Galaxy could have interesting observational implications. For instance, single or binary IMBHs might be detectable through microlensing \citep{paczynski_1986}. IMBHs ejected from the cluster could capture luminous companions, allowing astrometric measurements to reveal their presence in the binary, as demonstrated by Gaia’s discovery of BH–star systems \citep{elbadry_gaiabha, elbadry_gaiaBHb, chak_gaiabh_2023, tanikawa2024_gaiabh, gaia_bh3}. Furthermore, ejected IMBH–BH binaries may emit low-frequency gravitational waves detectable by interferometers such as LISA \citep[see e.g.][]{arcasedda_2021}.
\begin{table*}[]
\centering
\setlength{\tabcolsep}{5pt}
\small
\begin{tabular}{lccccc}
\toprule
Name & $\log (M_{\rm cl}/\msun)$ & $\sigma_{\log M_{\rm cl}}$ & $\log (M_{\rm IMBH}/\msun)$ & $\sigma_{\log M_{\rm IMBH}}$ & $\log \mathcal{B}^{\rm N}_{\rm G}$ \\
\midrule
G1        & 6.76 & 0.02 & 4.25 & 0.11 & 2.35 \\
NGC5139 ($\omega$ Cen)\textsuperscript{\textdagger}  & 6.40 & 0.05 & $\sim$ (3.77 - 3.91) & - & 2.04 \\
NGC6715 (M54)   & 6.28 & 0.05 & 3.97 & 0.18 & 0.77 \\
NGC6388   & 6.04 & 0.08 & 4.23 & 0.18 & 0.21 \\
NGC6266 (M62)   & 5.97 & 0.01 & 3.30 & 0.18 & 0.13 \\
NGC104 (47 Tuc)\textsuperscript{\textdagger}\textsuperscript{\textdagger}  & 5.88 & 0.08 & 3.36 & (0.16, 0.28) & 0.13 \\
NGC2808   & 5.91 & 0.04 & 4.00 & - & 0.03 \\
NGC7078 (M15)  & 5.79 & 0.02 & 3.64 & - & -0.15 \\
NGC6093 (M80)  & 5.53 & 0.03 & 2.90 & - & -0.27 \\
NGC5824   & 5.65 & 0.03 & 3.78 & - & -0.29 \\
NGC1851   & 5.57 & 0.04 & 3.30 & - & -0.37 \\
NGC5286   & 5.45 & 0.02 & 3.17 & 0.24 & -0.47 \\
NGC5694   & 5.41 & 0.05 & 3.90 & - & -0.51 \\
NGC1904 (M79)  & 5.15 & 0.03 & 3.47 & 0.12 & -0.77 \\
\bottomrule
\end{tabular}
 \caption{\small Milky Way \globular potentially hosting an IMBH and G1 cluster in the Andromeda galaxy. Column 1 refers to the nomenclature of each cluster, Column 2 and 3 to the logarithm of the cluster mass and its uncertainty, Column 4 and 5 the same as 2 and 3 but for the logarithm of the IMBH mass. Finally, Column 6 presents the logarithm of the Bayes factor evaluated as in Eq.~\eqref{eq::bayesfactor}. For all the clusters we consider the values in \cite{lutz2013_gcs}, with the exception of \textsuperscript{\textdagger} \citep{haberle_omegacen_2024, banares_2025} and \textsuperscript{\textdagger}\textsuperscript{\textdagger} \citep{kiziltan_2017}. }
\label{tab::bayes}
\end{table*}

\begin{figure}[h!]
    \centering
    \begin{subfigure}[t]{0.8\columnwidth}
        \includegraphics[width=\textwidth]{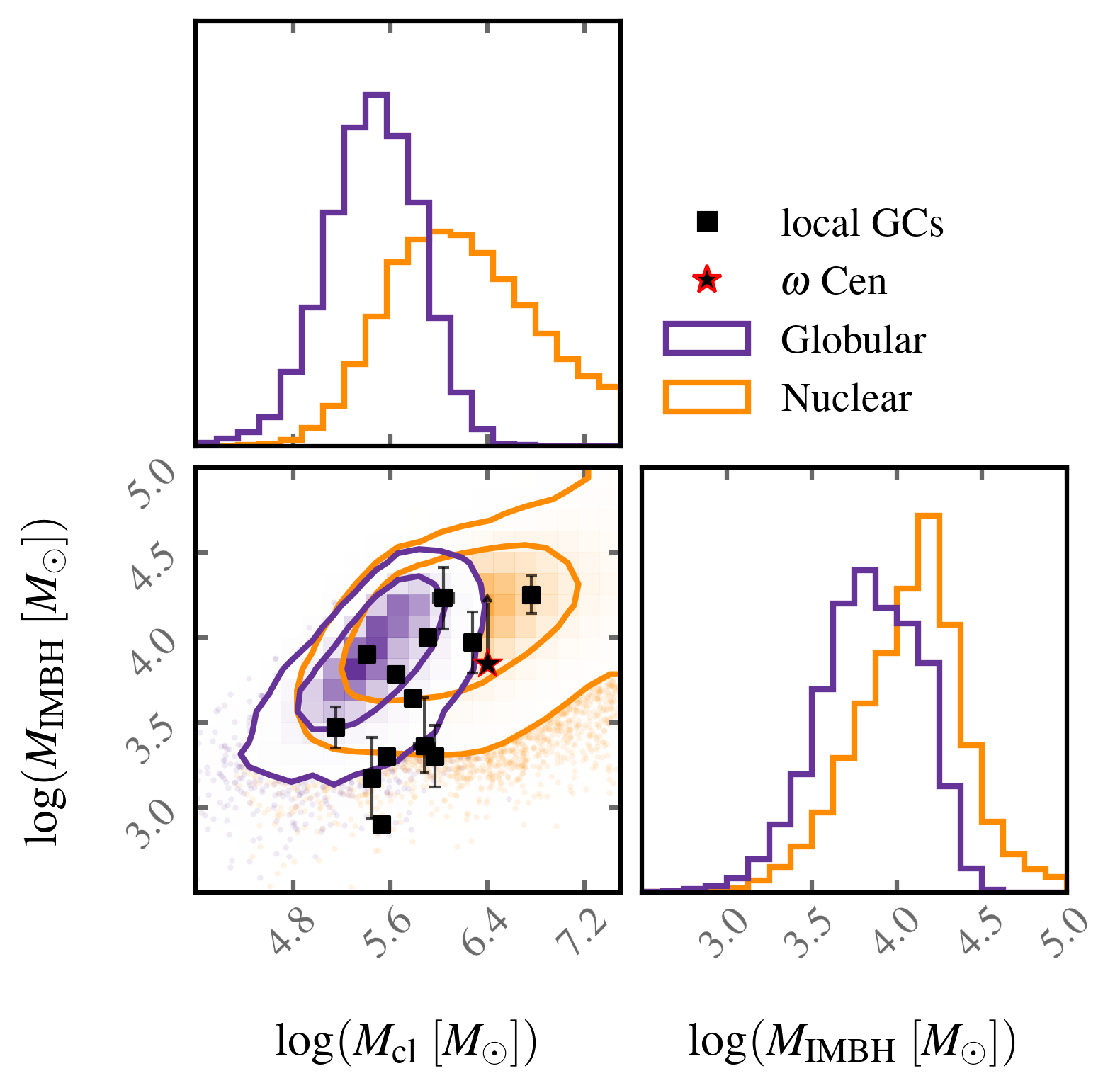}
        \caption{}
        \label{fig:corner_mwcandidates_a}
    \end{subfigure}
    
    \vspace{0.5cm} 
    
    \begin{subfigure}[t]{0.8\columnwidth}
        \includegraphics[width=\textwidth]{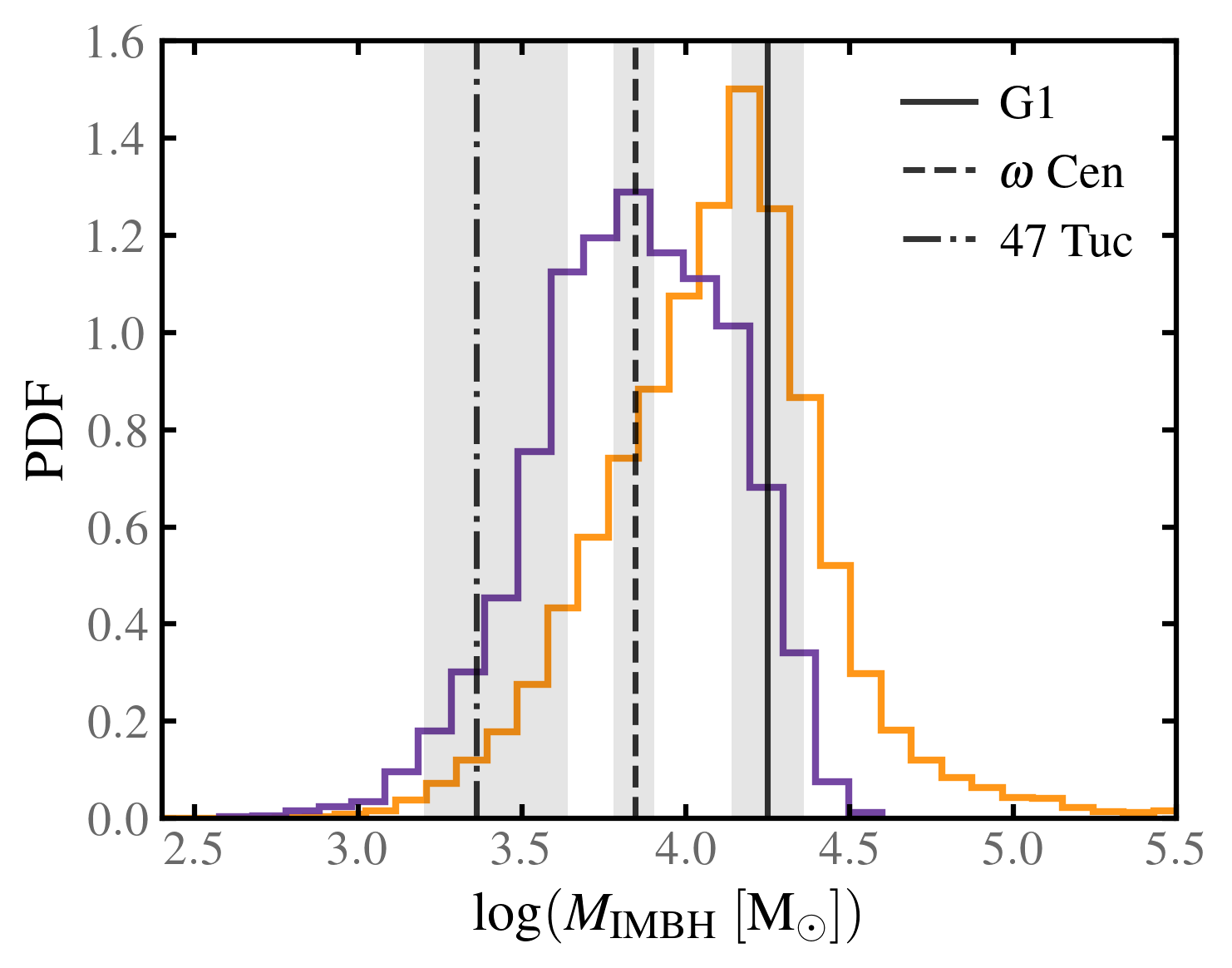}
        \caption{}
        \label{fig:corner_mwcandidates_b}
    \end{subfigure}
    
    \caption{\small Comparison of simulated IMBH candidates in Model B with observational data. \textbf{(a)} Corner plot showing the \globular (purple) and \nuclear (orange) clusters in our simulations hosting an IMBH at $z=0$, compared to observations of potential IMBHs in Milky Way globular clusters and G1 \textbf{(b)} One-dimensional IMBH mass distribution with comparison to estimated masses for G1 and 47 Tuc.
    }
    \label{fig:combined_mwcandidates}
\end{figure}

\subsection{Galactic globular clusters hosting IMBHs and their (mis)classification}
\label{sec::bayes}
As discussed in Section \ref{subsec::lowredcandidates}, stellar collisions can explain potential IMBH observations at $z = 0$ in clusters with masses below $\sim 10^7 \, \msun$. However, the agreement between our simulations and candidate \globular hosts \citep{lutz2013_gcs, kiziltan_2017, haberle_omegacen_2024} depends strongly on the assumed IMBH seeding model. Figure \ref{fig::allIMBHs} shows that a runaway stellar collision scenario (Model A) struggles to reproduce these observations, whereas a milder stellar collision scenario (Model B) aligns significantly better with the data.

Focusing on Model B, we find that both \globular\ and \nuclear can partially recover the same host–IMBH masses inferred for these observations (Figure \ref{fig:corner_mwcandidates_a}). To quantify this, we introduce a simple Bayesian framework to evaluate whether local IMBH host candidates are more likely \globular or \nuclear, offering a systematic approach for studying the Milky Way’s \globular.
This is particularly relevant since the Galactic population of \globular\ is expected to be mixed. Theoretical models and observations suggest indeed that a fraction of the Milky Way’s \globular\ formed in situ, while $\sim 30-35\%$ were accreted during galaxy mergers \citep[see e.g.][]{massari_2019}, with some of the latter possibly being stripped galactic nuclei. Since our simulations reveal clear links between the properties of a cluster and its IMBH, they provide a powerful tool to differentiate between genuine \globular and former \nuclear among local IMBH hosts.

To summarize, our goal is to use the Bayes factor to compare two competing hypotheses for the nature of the observed IMBH hosts:  
\begin{enumerate}
    \item $H_{\rm GC}$: the observed IMBH host is a GC;  
    \\
    \item $H_{\rm NSC}$: the observed IMBH host is a NSC.
\end{enumerate}
As already mentioned, we restrict our analysis to Model B simulation (right panel of Figure \ref{fig::allIMBHs}), which shows the most significant overlap with the observed candidates.
From this simulation we obtain discrete samples of the logarithm of the intrinsic cluster mass and IMBH mass, i.e. $\{\log(M_{\rm cl}), \log(M_{\rm IMBH})\}$. To convert these samples into continuous probability densities we smooth them using a gaussian kernel density estimator (KDE). 
 Figure~\ref{fig:corner_mwcandidates_a} shows a zoom-in of these simulated distributions together with the inferred cluster and IMBH masses of a set of known Milky Way and local \globular host candidates (black squares). These masses are also listed in Table~\ref{tab::bayes}. We gather these estimates from a collection of different studies \citep{lutz2013_gcs, kiziltan_2017, haberle_omegacen_2024, hernandez_omegacen_2025} as follows. 
 For the clusters with only upper limits on IMBH masses in \cite{lutz2013_gcs}, we conservatively treat these limits as representative values. For $\omega$ Centauri ($\omega$ Cen), we rely instead on two recent and somewhat conflicting analyses. \citet{haberle_omegacen_2024} infer a lower limit of $\sim 8200 \, \msun$ based on stellar kinematics, whereas \citet{hernandez_omegacen_2025} favor an extended central mass of $\sim 2-3 \times 10^5 \, \msun$, setting a lower limit of $\sim 6000 \, \msun$ for an IMBH. To bracket these uncertainties, we model the IMBH mass in $\omega$ Cen as uniformly distributed between 6000 and 8200 $\msun$. Finally, for 47 Tucanae (47 Tuc) we adopt the cluster and IMBH mass estimates reported by \citet{kiziltan_2017}. 
For each set of inferred masses, we then assign a posterior probability distribution function, which we approximate as a two-dimensional Gaussian with means and standard deviations given by the reference values in Table~\ref{tab::bayes}. 

In Appendix \ref{app::bayes} we derive an approximated expression of the Bayes factor relative to a single observation assuming that the likelihood is effectively the same for $H_{\rm GC}$ and $H_{\rm NSC}$, and exploiting the fact that the likelihood can be derived directly from the posterior of $\{\log(M_{\rm cl}), \log(M_{\rm IMBH})\}$ if the prior is sufficiently broad and uniform. 
The values of the Bayes factors' logarithms for each host candidate are listed in the last column of Table \ref{tab::bayes} as $\log \mathcal{B}^{\rm N}_{\rm G}$. 
Note that the rows of Table \ref{tab::bayes} are ordered according to $\log \mathcal{B}^{\rm N}_{\rm G}$ in descending order, with
positive (negative) values supporting a NSC (GC) hypothesis. In particular, we can consider $\log \mathcal{B}^N_G > 2$ ($< -2$) as strongly pointing towards a NSC (GC) nature of the cluster \citep{jeffreys_1939}. Two notable cases which emerge from our simple analysis are the G1 cluster, which orbits the Andromeda galaxy, and $\omega$ Cen in the Milky Way, both showing substantial preference for a NSC origin. G1 is hypothesized to be the disrupted nucleus of a dwarf galaxy that merged with Andromeda, a scenario supported by its broad stellar metallicity distribution \citep[see e.g.][]{g1_2007}. Similarly, $\omega$ Cen, the most massive cluster in the Milky Way, is also one possible example of an accreted nuclei \citep{omegacen_hilker2000, omegacen_ibata2019}, though the presence of an IMBH in this cluster has long been debated. Most clusters, like 47 Tuc, show a a slight or negligible preference to the NSC or GC hypothesis. For reference, in Figure \ref{fig:corner_mwcandidates_b}, we also present a zoom-in of the IMBH mass distributions in \globular and \nuclear and the putative masses of the IMBHs contained in G1, $\omega$ Cen and 47 Tuc. Note that, while G1 falls on the peak of the \nuclear distribution found in our simulations, the $\omega$ Cen IMBH mass is more compatible with the IMBHs found in \globular. Nevertheless the large mass of this cluster ultimately determines its preference towards a NSC origin. 

\section{Conclusions}
\label{sec:conclusions}
We have investigated the formation and growth mechanisms of IMBHs in young (\young), globular (\globular) and nuclear (\nuclear) clusters by means of the new semi-analytic code \bpop (\citealp{arcasedda_benacquista_2019, arcasedda_isolated_2023}; Arca Sedda et al., in prep.). 
We focused on the potential formation of IMBHs through two mutually non-exclusive channels, namely  (i) the collapse of very massive stars (VMSs) resulting from repeated stellar collisions and  (ii) hierarchical BBH mergers. For the first channel, we adopted two models: Model A, assuming VMS formation via runaway stellar collisions in clusters with short core-collapse times ($t_{\rm cc} < 5\,\mathrm{Myr}$; \citealt{portegies_zwart_runaway_2002}), and Model B, invoking a mild collision scenario in which multiple stellar collisions in extremely dense clusters drive the VMSs formation ($\rho_{\rm cl,0} > 10^{5}\,\mathrm{M}_\odot\,\mathrm{pc}^{-3}$) and motivated by recent $N$-body simulations \citep{dragonII_II_2023, rantala2025rapidchannelcollisionalformation}. 
In both cases we restricted the seeding to metal-poor environments ($Z < 0.001$) to avoid strong stellar-wind mass loss.
Furthermore, we tested a separate scenario using the same seeding as Model A but with alternative redshift distributions for \nuclear\ and \globular (Model A'), to examine the effects of clusters formation histories on IMBH assembly.

Our main results can be summarised as follows:
\begin{itemize}
\item The seeding conditions and cluster formation histories strongly shape the IMBH formation efficiency across cluster families. The runaway stellar collision scenario (Model A), which favors systems with short core-collapse times, maximizes the efficiency in \globular. Conversely, the mild collision scenario (Model B) is the most efficient in \nuclear, which comprise the densest clusters. In Model A' \nuclear prove to be much more efficient at producing IMBH seeds compared to \globular, owing to their broader distributions in redshift and metallicity, which extends down to $Z < 0.001$. Hierarchical IMBHs form efficiently only in \nuclear and always with lower efficiency than IMBH seeds. As a result, the majority of IMRIs originate from IMBH seeds rather than from hierarchically assembled IMBHs.
\\

\item The sub-population of IMBHs retained in their host clusters at $z = 0$ in Model A exhibits a distinct gap in the cluster mass - IMBH mass plane around $M_{\rm cl} \sim 10^7\,\msun$. This scarcity of IMBH hosts results from the combined effects of 
the seeding conditions and the inefficiency of the hierarchical channel IMBH formation channel. In contrast, Model B presents an homogeneous population of retained IMBHs which smoothly transitions from IMBHs seeded by stellar collisions to those assembled hierarchically via BBH mergers. These IMBHs have a clear overlap with potential observations of IMBHs in galactic \globular.
\end{itemize}
Additionally, we discuss possible implications for IMBHs in \young and \globular:
\begin{itemize}
    \item We show that if stellar collisions are an efficient mechanism for IMBH formation, we likely expect a sub-population of ejected IMBH seeds wandering in the potential well of Milky Way-like galaxies. The presence of such IMBHs (as single objects or in binaries) could be probed through different observational channels, such as astrometric measurements of a potential stellar companion (similarly to Gaia-BH1, Gaia-BH2, and Gaia-BH3 \citealt{elbadry_gaiabha, elbadry_gaiaBHb, chak_gaiabh_2023, tanikawa2024_gaiabh, gaia_bh3}), microlensing \citep{paczynski_1986} or GW emission signatures \citep{arcasedda_2021}.
    \\
    \item We discuss a simple Bayesian framework which exploits the correlations found in our simulations between the clusters and IMBHs masses to probe the nature of Milky Way IMBH host candidates. This is particularly interesting as some of the Milky Way \globular, e.g. $\omega$ Cen, are believed to actually be stripped nuclei of galaxies which interacted with our galaxy in the past. 
    We apply this approach comparing the systems found in Model B to the inferred masses in \cite{lutz2013_gcs, kiziltan_2017, haberle_omegacen_2024, hernandez_omegacen_2025}. 
    We find that most clusters lie in a region where the \globular and \nuclear hypotheses are equally probable with the exception of the G1 cluster in the Andromeda galaxy and $\omega$ Cen which strongly support a NSC origin.

\end{itemize}
Overall, our work highlights the critical role of stellar collisions in producing IMBHs over a broad range of clusters. Moreover, we propose possible ways of probing the presence of such IMBHs as well as disentangling their formation mechanisms and the formation histories of their hosts.
We also emphasize the fundamental importance of simultaneously considering the codependencies among different formation mechanisms, cluster families and their evolutionary histories, as well as environmental factors, to effectively investigate IMBH production and growth. In this context, the \bpop code represents a valuable tool for this kind of studies, as it integrates all these aspects within a flexible and efficient framework. 

\begin{acknowledgements}
    The authors acknowledge Gor Oganesyan, Filippo Santoliquido, and Konstantinos Kritos for useful discussion and comments. 
    MAS acknowledges funding from the European Union’s Horizon 2020 research and innovation programme under the Marie Skłodowska-Curie grant agreement No.~101025436 (project GRACE-BH). MAS and CU acknowledge financial support from the MERAC Foundation under the 2023 MERAC prize and support to research.  
    This research made use of \textsc{NumPy} \citep{harris2020}, \textsc{SciPy} \citep{scipy2020}, \textsc{Pandas} \citep{pandas2020}. For the plots we used \textsc{Matplotlib} \citep{hunter2007}.
\end{acknowledgements}

\bibliography{biblio}

\appendix

\section{Depletion of massive stars due to stellar collisions and minimum cluster mass}
\label{app::depletion}
Stellar collisions typically involve the most massive stars in clusters, both because of their large collision cross sections and because of their overabundance in the cluster core due to mass segregation. Consequently, stellar collisions can deplete the cluster of potential stellar-mass BH progenitors, potentially impacting the subsequent growth of IMBH seeds via BBH mergers. In this section, we show that this depletion process does not significantly affect the population of stellar-mass BH progenitors at the low metallicities required to sustain the growth of VMSs through stellar collisions.

Let us consider a Kroupa-like initial mass function (IMF) \citep{kroupa_2001} defined as:
\begin{equation}
    \xi(m) := \frac{dN}{dm} = 
    \begin{cases}
        A \times m^{-\alpha} & m \in [0.08, 0.5]\,\msun \\[6pt]
        B \times m^{-\beta} & m \in [0.5, 150]\,\msun
    \end{cases}
\end{equation}
with $\alpha = 1.35$, $\beta = 2.35$, and normalization constants $A, B$ chosen such that $\int \xi(m) \times m \, dm = 1$ and $\xi(m)$ is continuous at $m = 0.5\,\msun$.
Assuming that stellar collisions preferentially consume stars from the heaviest to the lightest, we estimate the minimum mass cutoff $m_{\rm cut}$ such that the fraction of stellar mass above $m_{\rm cut}$ equals the fraction of mass accreted through stellar collisions. While this is a simplification, since collisions are stochastic and may involve lighter stars, it allows us to identify the stellar mass range that primarily contributes to the final VMS mass. 
The fraction of mass accreted by a seed star via stellar collisions can be simply expressed as:
\begin{equation}
    f_{\rm *} = \frac{m_{\rm VMS} - m_{\rm seed}}{M_{\rm cl}} \ , 
\end{equation}
with $m_{\rm VMS}$ ($m_{\rm seed}$) the final (seed) mass of the VMS. In a runaway stellar collisions scenario (Model A) this fraction can be computed using Eq.~\eqref{eq::mvms_simplified} and corresponds to a value $f_{\rm *, A} \sim 0.008$. Equating $f_{\rm *, A}$ to the integral of the high-mass tail of the IMF, 
\begin{equation}
    \int_{m_{\rm cut}}^{150\,\msun} \xi(m) \times m \, dm \ ,
    \label{eq::appIMFintegral}
\end{equation}
we find $m_{\rm cut} \sim 110\,\msun$. For metallicities $Z < 10^{-3}$, this mass lies near the low-mass edge of the pair-instability supernovae (PISNe) gap (see \citealp{sevn_2017, iorio_2023} for reference), although the exact gap boundaries depend on stellar evolution models. Therefore, we reasonably neglect depletion in our simulations, assuming that most of the VMS final mass originates from stars that would otherwise have undergone PISNe.

Similarly, for the mild collisions seeding scenario (Model B), we can estimate the maximum expected mass fraction from $m_{\rm VMS, max}$ in  Eq.~\eqref{eq::mvms_bifrost}, $f_{\rm *, B} \sim 0.02$, and find a cutoff mass from Eq. \eqref{eq::appIMFintegral} $m_{\rm cut} \sim 93\,\msun$, which indicates a mild depletion of massive stars. Note, however, that IMBH seeds typically do not merge with all available BHs in the cluster, hence we expect that neglecting the depletion in this case would still not affect significantly our final results.

Our calculations implicitly assume that the cluster stellar population closely follows the theoretical IMF. However, if the cluster contains too few stars, the high-mass end of the IMF will be poorly sampled. To estimate the minimum cluster mass required to host at least one star heavier than $m_{\rm cut}$, we compute the average number of stars in the cluster as $N_* = \frac{M_{\rm cl}}{\langle m_{\rm Kroupa} \rangle} \approx \frac{M_{\rm cl}}{0.7\,\msun}$, and solve for $M_{\rm cl}$ such that $N_* \times \int_{m_{\rm cut}}^{150\,\msun} \xi(m) \, dm = 1$
. This yields a minimum cluster mass of approximately $2 \times 10^4\,\msun$ for the runaway seeding scenario (Model A) and $5 \times 10^3\,\msun$ for the alternative seeding scenario (Model B). Accordingly, we allow seeding only in clusters heavier than these thresholds for the respective models.

\section{BH dynamics in B-POP}
\label{app::theory}
In the following we provide an overview of the modeling of BH dynamics in \bpop. We refer to our companion paper (Arca Sedda et al., in prep.) for a detailed description of the clusters mass and half-mass radius evolution modeling.

Given a cluster of initial mass $M_{\rm cl,0}$, \bpop firstly sets up the number of possibly interacting BHs per cluster ($n_{\rm BHs}$) assuming it to be comparable to the number of BHs inside the cluster scale radius, i.e.  
\begin{equation}
    n_{\rm BHs} = f_{\rm BH} \ f_{\rm reten} \ f_{\rm encl}\ M_{\rm cl,0} \ ,
    \label{eq::nbhs}
\end{equation}
with $f_{\rm BH}= 0.0008$ the fraction of cluster mass in stellar BHs according to a \cite{kroupa_2001} IMF, $f_{\rm reten}= 0.5$ the fraction of BHs retained in the cluster, and $f_{\rm encl}$ the fraction of cluster mass enclosed within the cluster typical radius.

For each cluster the code follows the evolution of a single BH (potentially an IMBH seed) and assumes the reservoir of surrounding BHs to be $n_{\rm BHs}$.
The initial absolute time considered by the code is the formation time of the cluster which varies depending on the cluster type and the assumed formation history (see also Section \ref{subsec::models}). To this timescale \bpop adds up the BH stellar progenitor lifetime, taken from the input stellar catalogues. Note that the BH can be already in a binary with a certain probability which is specified as one of \bpop input parameters. Afterwards, \bpop considers the BH or BBH segregation timescale to the center of the cluster which was already introduced in Eq. \eqref{eq::tseg}.
If the BH is not part of a binary, once it segregates to the core it can pair up with another BH via either 3-body or binary-single scatterings.
In the first scenario the BBH forms as a result of the dynamical interaction of three unbound objects over a timescale \citep{lee_1993}:
\begin{multline}
    t_{\rm 3bb} = 4 \, \mathrm{Gyr}\,
    \left( \frac{10^6 \, \msun\,\mathrm{pc}^{-3}}{\rho_{\rm c}} \right)\, 
    \left( \zeta^{-1} \frac{\sigma_{\rm c}}{30\,\mathrm{km/s}} \right)^9     \times \\
        \vspace{0.5cm}
    \left( \frac{10 \, m_*}{m}\right)^{9/2}\,\left( \frac{10\,\msun}{m} \right)^{-5}\,,
    \label{eq::appt3bb}
\end{multline}
where $\rho_{\rm c}$ and $\sigma_{\rm c}$ are the core density and velocity dispersion, while $\zeta \leq 1$ parametrizes the level of energy equipartition between the heavy and light populations of stars in the cluster.\\
In the second scenario the BH is captured by a pre-existing stellar binary on a timescale \citep{Miller_2009}:
\begin{multline}
    t_{\rm bs} = 3 \, \mathrm{Gyr} \, \left( \frac{0.01}{f_{\rm b}}\right) \, \left( \frac{10^6\, \mathrm{pc}^{-3}}{n_*}\right) \times\\
    \vspace{0.5cm}
        \left( \frac{\sigma_{\rm c}}{30 \, \rm km/s} \right) \, \left( \frac{10 \, \msun}{a_{\rm h} (m + m_1 + m_2)}\right)\,.
    \label{eq::apptbs}
\end{multline}
In Eq. \eqref{eq::apptbs}, $m_1$ and $m_2$ correspond to the masses of the two stars, $f_{\rm b}$ to the fraction of primordial stellar binaries in the cluster, $n_*$ to the star number density, while the orbital separation:
$$a_{\rm h} \sim 59\,\mathrm{a.u.}\,\left(\frac{m_1 + m_2}{30 \, \msun} \right) \ \left(\frac{30\,\mathrm{km/s} }{\sigma_{\rm c}}\right)$$ represents the typical hard binary separation.

A hard binary is characterized by a binding energy greater than the typical kinetic energy of stars in the cluster ($G\,m_1 m_2/a > \,m_*\,\sigma^2$). On the contrary, a binary with a binding energy smaller than the average stellar kinetic energy is referred as to a soft binary.

A hard binary will continue to harden (to get tigher) on average as a result of close dynamical interactions with the other objects in the cluster, whereas a soft binary will continue to soften \citep{heggie_binary_1975}. 
BHs are tipically in hard binaries since they are among the most massive objects in a cluster.

Though dynamical interactions favor the shrinking of the orbital separation they can also result in the premature expulsion of the BBH from the cluster. Indeed, these interactions also give off some energy to the binary center of mass, producing a recoil which gets increasingly stronger as the binary gets harder. 
In \bpop we evaluate the critical orbital distance at which the binary recoil exceeds the escape velocity of the cluster and ejects the binary as \citep{antoninirasio2016}:
\begin{equation}
    a_{\rm ej} = 0.07\,\mathrm{a.u.} \, \frac{\mu\,m_{\rm p}^2}{(m_1 + m_2 + m_{\rm p})\,(m_1+m_2)}\, \left(\frac{v_{\rm esc}}{50\,\mathrm{km/s}}\right)^{-2}\,,
    \label{eq::appaej}
\end{equation}
where $m_1$, $m_2$ are the masses of the BHs in the binary, $\mu = m_1 m_2 / (m_1+m_2)$ is the binary reduced mass, and $m_{\rm p}$ is the average mass of a BH perturber.
This dynamical hardening competes with the hardening driven by GW emission which becomes efficient at an orbital separation:
\begin{multline}
        a_{\rm GW} = 0.05\,\mathrm{a.u.} \, \left( \frac{m_1 + m_2}{ 20\,\msun }\right)^{3/5}\,\left( \frac{(m_2/m_1)^{1/5}}{(1+m_2/m_1)^{2/5} }\right) \times \\
        \vspace{0.5cm}
        \left( \frac{\sigma_{\rm c}}{ 30\,\mathrm{km/s} }\right)^{1/5}\,\left( \frac{10^6\,\msun\,\mathrm{pc^{-3}}}{ \rho_{\rm c}}\right)^{1/5} \
        \label{eq::appagw}
\end{multline}
The code computes the orbital separations in Eq. \eqref{eq::appagw} and \eqref{eq::appaej} and considers the maximum of the two as the critical separation orbit, i.e. $a_{\rm crit} = \max(a_{\rm ej}, a_{\rm GW}$). This orbital separation is assumed to be reached after a timescale \citep{millerhamilton2002}:
\begin{equation}
    t_{\rm crit} \sim 5 \, \frac{m_1 + m_2}{m_{\rm p}} \ t_{2-1}\,,
    \label{eq::apptmercrit}
\end{equation}
with $t_{2-1}$ the typical timescale over which the binary hardens due to single-binary interactions which is evaluated in \bpop as \citep{gultekin_2004, antoninirasio2016}:
\begin{multline}
    t_{2-1} = 0.02 \, \mathrm{Gyr} \, \zeta^{-1} \, \left( \frac{10^6 \, \mathrm{pc}^{-3}}{n_*}\right) \, \left( \frac{\sigma_{\rm c}}{\mathrm{30 \, km/s}}\right) \times \\
    \left( \frac{0.05\,\mathrm{a.u.}}{a_{\rm h}}\right)\,\left( \frac{20\,\msun}{m_1 + m_2}\right)\,\sqrt{\frac{10\,m_*}{m_{\rm p}}}\,.
    \label{eq::appt12crit}
\end{multline}
If $a_{\rm crit} = a_{\rm ej}$ \bpop re-samples the BBH eccentricity from a thermal distribution and checks if $t_{\rm crit}$ is smaller than the GW merger time ($t_{\rm GW}$) evaluated as in \cite{peters_matthews_1963}. If $t_{\rm GW} < t_{\rm crit}$ the code assumes the binary to merge within the cluster on a merger time $t_{\rm mer} = t_{\rm GW}$. This corresponds to a scenario in which the BBH merger is caused by a sudden eccentricity change induced by strong dynamical interactions. If $t_{\rm GW} > t_{\rm crit}$ the binary is considered to be ejected and to merge over a timescale $t_{\rm crit} + t_{\rm GW}$. Conversely, whenever $a_{\rm crit} = a_{\rm GW}$ \bpop assumes the BBH to merge within the cluster on a timescale $t_{\rm crit} + t_{\rm GW}$. 
If the total time needed to reach the merger is longer than a Hubble time the binary is considered to not be merging. When the BBH successfully merges within the host cluster the code evaluates the relativistic kick induced on the remnant BH and compares it to the cluster escape velocity at the moment of the merger.
If the kick does not exceed the cluster escape velocity, the BH can undergo subsequent cycles of evolution and experience multiple mergers, until it is either (i) ejected by a Newtonian recoil in a binary interaction or by a post-merger relativistic kick as a single object, (ii) retained in a binary that does not merge within a Hubble time, or (iii) left without companions after exhausting all other BHs in the cluster.

\section{Gap dependence on the initial conditions}
\label{app:gap_position}
\begin{figure}[t]
    \centering
    \includegraphics[width=0.75\columnwidth]{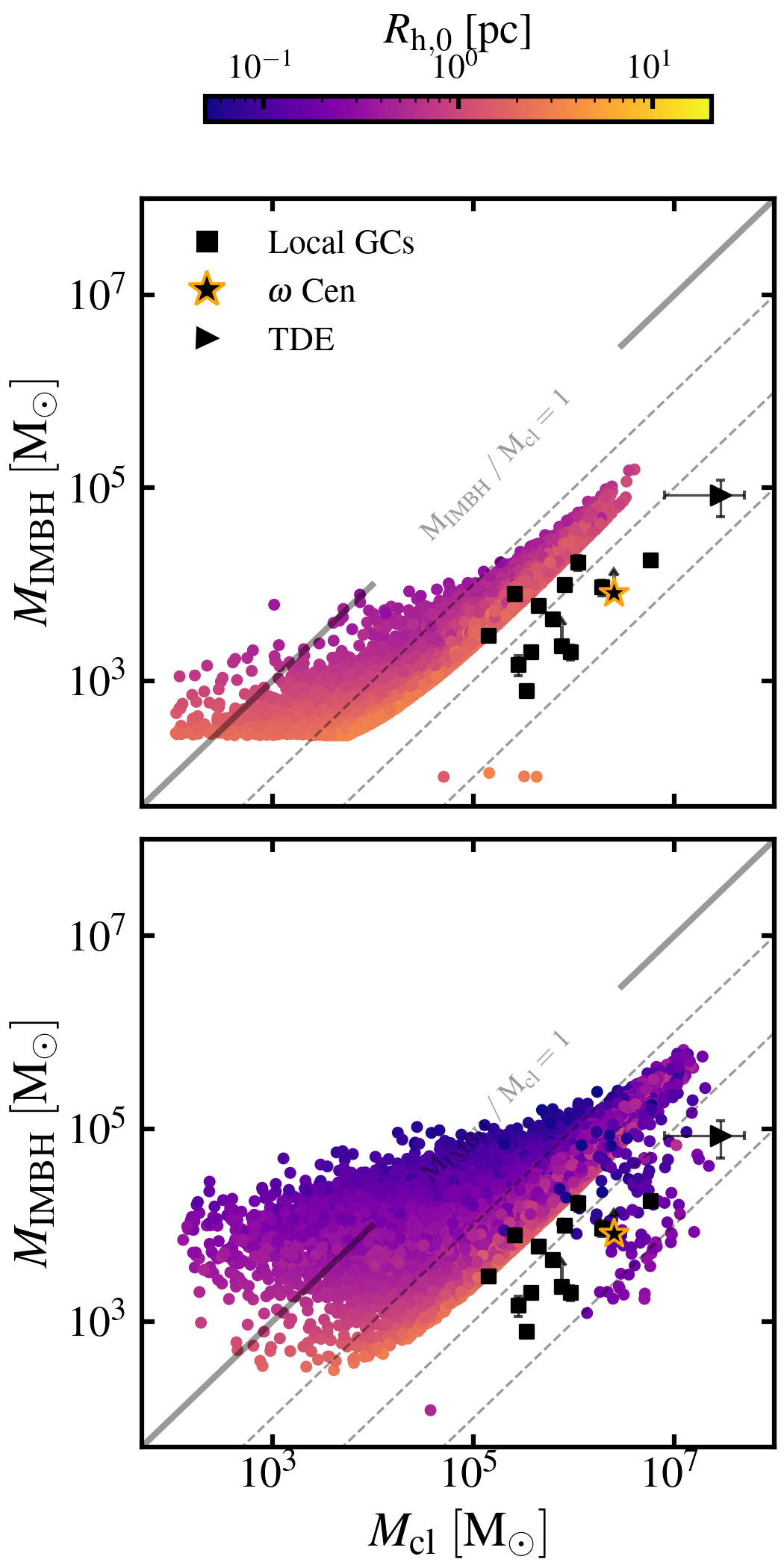}
\caption{
\small \globular hosting an IMBH at $z = 0$, compared to observed host candidates in local \globular \citep[squares;][]{lutz2013_gcs} and a potential TDE \citep[right triangle;][]{tde_lin_2018}, for two sets of initial mass--half-mass radius conditions. The orange-edged star marks the $\omega$ Cen cluster \citep{haberle_omegacen_2024}. As in Figure \ref{fig::allIMBHs}, the dashed lines correspond to mass ratios $M_{\rm IMBH}/ M_{\rm cl} = 10^{-1}, 10^{-2}, 10^{-3}$. The top panel shows results from a \bpop simulation of $10^8$ \globular, while the bottom panel shows results from a simulation in which the same \globular have initial masses larger by 0.5 dex and half-mass radii smaller by 1 dex.
}
\label{fig:allIMBHs_enhanced}

\end{figure}

As discussed in Section \ref{subsec::lowredcandidates}, we observe a dearth of retained IMBHs at $z = 0$ for clusters with masses around $\sim 10^7 \, \msun$ when assuming the seeding to be efficient only for clusters with very short core-collapse times, namely $<5\,\rm Myr$ (Model A). This gap in the cluster mass - IMBH mass plane, which we do not observe in Model B, is caused by the combined effect of the seeding condition and the initial mass - half-mass radius distributions assumed for the clusters. To assess the effect of assuming different initial conditions on the gap we run a separate simulation of $10^8$ BHs in \globular where we assume (i) the masses of the clusters to be enhanced by 0.5 dex and (ii) the half-mass radii to be decreased of 1 dex. These conditions are chosen to artificially enhance the number of clusters in the runaway seeding regime (see Eq. \eqref{eq::tseg} and \eqref{eq::trel}).
In Figure \ref{fig:allIMBHs_enhanced} we compare the obtained population of retained IMBHs to the one obtained for IMBHs in \globular in Model A (see Figure \ref{fig::allIMBHs}). As expected, we find an increase of IMBH seeds hosts. We also observe an increase of systems above the $M_{\rm IMBH} / M_{\rm cl} = 1 $ contour, since the overall decrease in the \globular relaxation times corresponds to higher mass losses over the simulation time. Interestingly, we also observe an enhanced formation of IMBHs through hierarchical mergers of stellar-mass BHs, corresponding to the points lying below the tail of seeded IMBHs and mainly clustered at $M_{\rm cl} \gtrsim 10^5 \, \msun$. These IMBHs are more easily produced under these initial conditions thanks to the increase in the escape velocity of \globular, which is of the order of $\sim \sqrt{M_{\rm cl, 0} / R_{\rm h, 0} } = 0.75 \rm \ dex \sim 5.6$. These IMBHs mildly overlap with galactic \globular host candidates though their abundance is still very limited with respect to the systems found in Model B (right panel of Figure \ref{fig::allIMBHs}).

\section{Bayes factor for cluster classification}
\label{app::bayes}
In Section~\ref{sec::bayes} we employed the Bayes factor to assess the relative support for two competing hypotheses regarding the nature of some \globular IMBH host candidates \citep{lutz2013_gcs, kiziltan_2017,haberle_omegacen_2024, hernandez_omegacen_2025}: a globular cluster origin ($H_{\rm GC}$) or a nuclear star cluster origin ($H_{\rm NSC}$). The Bayes factor is defined as
\begin{equation}
    \mathcal{B}^{\rm N}_{\rm G} =
    \frac{\mathcal{Z}(\vec{d}\,|\,H_{\rm NSC})}{\mathcal{Z}(\vec{d}\,|\,H_{\rm GC})} \, ,
    \label{eq::bayes}
\end{equation}
where
\begin{equation}
    \mathcal{Z}(\vec{d}\,|\,H_{i}) =
    \int d\vec{\theta} \,
    \mathcal{L}(\vec{d}\,|\,\vec{\theta},H_{i})\,
    \pi(\vec{\theta}\,|\,H_{i}) \, , 
    \quad H_i \in \{H_{\rm GC}, H_{\rm NSC}\} ,
    \label{eq::likedH}
\end{equation}
denotes the marginal likelihood, or evidence, of the data $\vec{d}$ under hypothesis $H_i$. TThe evidence is the likelihood of the data $\vec{d}$ given the intrinsic parameters $\vec{\theta}$, averaged over their prior distribution $\pi(\vec{\theta}|H_i)$ under hypothesis $H_i$. In our case, $\pi(\vec{\theta}|H_i)$ corresponds to the two–dimensional distributions of cluster and IMBH masses obtained from Model B simulation (see Figure~\ref{fig:corner_mwcandidates_a}).\\  
Since our simulations do not provide direct access to $\mathcal{L}(\vec{d},|,\vec{\theta},H_i)$, we instead apply Bayes’ theorem to rewrite the likelihood in terms of the observed posterior probability distributions. These posteriors are approximated as Gaussians based on the values reported in Table~\ref{tab::bayes} and we can denote them as $p(\vec{\theta}\,|\,\vec{d},H_0)$, where $H_0$ represents the specific model assumptions adopted in each observational study. By Bayes’ theorem,
\begin{equation}
    \mathcal{L}(\vec{d}\,|\,\vec{\theta},H_0) =
    \frac{p(\vec{\theta}\,|\,\vec{d},H_0)\,
    \mathcal{Z}(\vec{d}\,|\,H_0)}
    {\pi(\vec{\theta}\,|\,H_0)} \, ,
    \label{eq:Ldtheta}
\end{equation}
with $\pi(\vec{\theta}\,|\,H_0)$ the prior adopted in the corresponding inference, and $\mathcal{Z}(\vec{d}\,|\,H_0)$ the associated evidence.\\
In general, the likelihood may depend explicitly on the model assumed, i.e. $\mathcal{L}(\vec{d}\,|\,\vec{\theta},H_0)\neq \mathcal{L}(\vec{d}\,|\,\vec{\theta},H_{\rm GC}) \neq \mathcal{L}(\vec{d}\,|\,\vec{\theta},H_{\rm NSC})$. In the present context, however, the clusters observables (e.g. velocity dispersions, luminosity profile) are governed by the same dynamical and stellar–evolutionary processes, regardless of their nature, e.g. globular or nuclear. Consequently, we can assume a common likelihood $\mathcal{L}(\vec{d}\,|\,\vec{\theta}) =  \mathcal{L}(\vec{d}\,|\,\vec{\theta},H_0)$, and attribute all model dependence to the prior $\pi(\vec{\theta}|H_i)$.
The prior $\pi(\vec{\theta}|H_0)$ entering Eq.~\eqref{eq:Ldtheta} varies across observational studies and it is often not explicitly specified. For tractability, we approximate it as broad and uniform in both $\log_{10}(M_{\rm cl})$ and $\log_{10}(M_{\rm IMBH})$. Hence, substituting Eq.~\eqref{eq:Ldtheta} into Eq. \eqref{eq::bayes}, both $\mathcal{Z}(\vec{d}\,|\,H_0)$ and $\pi(\vec{\theta}|H_0)$ cancel out and the Bayes factor reduces to:
\begin{equation}
    \mathcal{B}^{\rm N}_{\rm G}  \sim \frac{\int d \vec{\theta} \ p(\vec{\theta}| \vec{d}, H_0) \ \pi(\vec{\theta} | H_{\rm NSC})}{\int d \vec{\theta} \ p(\vec{\theta}| \vec{d}, H_0) \ \pi(\vec{\theta} | H_{\rm GC})}
   \,.
    \label{eq::bayesfactor}
\end{equation}
Moreover, since the priors are assumed to be broad, no relevant regions of parameter space are truncated, and the integrals can be evaluated over the full domain of $\pi(\vec{\theta}|H_i)$.
Specifically, we estimate the integrals in Eq. \eqref{eq::bayesfactor} using Monte Carlo summation as:
\begin{equation}
   \sum_j^{N_{\rm s}} \pi(\vec{\theta}_j | H_i)\,,\,\mathrm{where} \,\vec{\theta_j} \sim p(\vec{\theta}| \vec{d}, H_0)\, , 
\end{equation}
i.e. sampling $\vec{\theta}_j$ from the the host candidate properties posterior ($p(\vec{\theta}| \vec{d}, H_0)$) and computing the probability of such system emerging from our simulation as a NSC or GC ($\pi(\vec{\theta}|H_{\rm NSC})$ or $\pi(\vec{\theta}|H_{\rm GC})$).

To actually select one hypothesis over the other we would need to include in Eq. \eqref{eq::bayesfactor} an extra term, the so-called prior-odds of the two hypotheses, i.e. $\pi (H_{\rm NSC}) / \pi(H_{GC})$. Including this term we would get:
\begin{equation}
    \mathcal{O}^N_G = \frac{p(H_{\rm NSC}| \vec{d})}{p(H_{\rm GC}| \vec{d})} = \mathcal{B}^N_G \ \frac{\pi(H_{\rm NSC})}{\pi(H_{GC})}\, ,
    \label{eq::postodds}
\end{equation}
where $\mathcal{O}^N_G$ is called posterior-odds. In our context, the prior odds encode the initial degree of belief about the nature of the observed hosts and can vary substantially depending on the adopted criteria. For example, one could attempt to evaluate it starting from the number of mergers the Milky Way has experienced with dwarf galaxies (i.e. the number of potential accreted nuclei) and the number of in-situ \globular\ IMBH hosts. However, these two quantities are highly non-trivial to estimate. Moreover, the prior odds may differ for each candidate depending on properties such as metallicity spread, density profile, or galactic position.

For simplicity, in this work we focus on estimating only the Bayes factor (Eq.~\ref{eq::bayesfactor}), which measures the relative support for the globular vs. nuclear hypothesis given the data. This corresponds to evaluating the posterior odds under the assumption of equal prior probabilities for $H_{\rm NSC }$ and $H_{\rm GC}$.

\end{document}